\documentclass[11pt]{article}

\usepackage{color}
\usepackage{soul}

\usepackage[utf8]{inputenc}
\usepackage{comment}
\usepackage{amssymb}

\usepackage[numbers]{natbib}
\usepackage{hyperref}

\usepackage{geometry}
\usepackage{graphicx}
\usepackage{booktabs}
\usepackage{array}
\usepackage{paralist}
\usepackage{verbatim}
\usepackage{bm}
\usepackage{subfig}
\usepackage{amsmath}
\usepackage{graphicx}
\usepackage{booktabs} 
\usepackage{multirow}
\usepackage{array,booktabs,tabularx}
\newcolumntype{C}[1]{>{\centering\arraybackslash}m{#1}}
\newcolumntype{R}{>{\raggedleft\arraybackslash}p{0.18\linewidth}}
\newcolumntype{Q}{>{\raggedright\arraybackslash}p{0.42\linewidth}} 

\usepackage{fancyhdr}
\usepackage{sectsty}
\allsectionsfont{\sffamily\mdseries\upshape}
\usepackage[nottoc,notlof,notlot]{tocbibind}
\usepackage[titles,subfigure]{tocloft}

\numberwithin{equation}{section}

\title{Preserving Mass Shell Condition in the Stochastic Optimal Control Derivation of the Dirac Equation}

\author{Vasil Yordanov \\
	v.yordanov@phys.uni-sofia.bg \\
	\textit{Faculty of Physics, Sofia University,}\\
	\textit{5 James Bourchier blvd., 1164 Sofia, Bulgaria}
}

\begin{document}
\maketitle
\abstract
The Dirac equation, central to relativistic quantum mechanics, governs spin-$\frac{1}{2}$ particles and their antiparticles, with each spinor component satisfying a Klein-Gordon-type equation -- the quantum counterpart of the relativistic mass-shell condition.
In our prior work [V. Yordanov, Sci. Rep. 14, 6507 (2024)], we derived the Dirac equation using stochastic optimal control (SOC) theory by linearizing both the Lagrangian's kinetic term and the Hamilton-Jacobi-Bellman (HJB) equation, but that earlier linearized derivation did not preserve the relativistic mass-shell condition. 
Here, we formulate the SOC problem starting from a specified covariant Lagrangian for a single relativistic charged particle in an external electromagnetic field, in which the standard relativistic square-root kinetic term and the minimal electromagnetic coupling are retained in their original form.
In addition, the Lagrangian is supplemented by a covariant spin-field coupling term describing the interaction of the particle's intrinsic spin with the electromagnetic field.
Because this spin-field term is matrix-valued, it is incorporated through scalarization so that the HJB problem remains scalar.
To retain a scalar HJB formulation, the present construction is restricted to electromagnetic backgrounds for which \(\sigma^{\mu\nu}F_{\mu\nu}\) admits a spacetime-independent eigenspinor; the Dirac equation is then obtained in the corresponding fixed spin sector.
This yields a relativistically consistent SOC derivation in which the $\hbar \to 0$ limit of the HJB equation recovers the proper-time relativistic Hamilton--Jacobi equation, whose stationary form yields the classical mass-shell condition, while the stationary HJB equation yields the quantum-corrected mass-shell relation.
We illustrate the theory with stochastic simulations of the Dirac-Landau problem (electron in a uniform magnetic field): at the analytic optimal drift the average stochastic action attains a local minimum, and the average action components are consistent with the Dirac-Landau values within statistical uncertainty.

\section{Introduction}
The Dirac equation~\cite{Dirac1928}:
\begin{equation}
\label{eq:dirac}
\gamma^\mu (i \hbar \partial_\mu - e A_\mu) \psi = m c  \psi,
\end{equation}
is a fundamental equation of relativistic quantum mechanics, elegantly unifies special relativity and quantum theory to describe spin-$\frac{1}{2}$ particles such as electrons. Its defining four-component spinor structure ensures each component obeys the Klein-Gordon equation, thereby enforcing the relativistic mass-shell condition ($E^2 = p^2 c^2 + m^2 c^4$). This feature guarantees that solutions remain consistent with special relativity's energy-momentum relation, essential for modeling relativistic quantum systems.

Stochastic mechanics, initiated by Fényes~\cite{Fenyes1952} and Nelson~\cite{Nelson1966}, reinterprets quantum mechanics via stochastic processes, yet extending it to derive the Dirac equation for spin-$\frac{1}{2}$ particles remains elusive~\cite{Blaquiere1979,Papiez1981,Gaveau1984,Kuipers2021,Lindgren2019,Simulik2025}. Past efforts often yielded the Klein-Gordon equation for spin-0 particles, missing the spinor structure of the Dirac equation.

In our earlier study~\cite{Yordanov2024}, we employed Stochastic Optimal Control (SOC) theory to derive the Dirac equation by linearizing both the Lagrangian's kinetic term, $-m c \sqrt{u_\mu u^\mu}$, and the second-order Hamilton-Jacobi-Bellman (HJB) equation.  
The resulting linearized HJB equation took the following factored form:
\begin{equation}
\label{eq:factored_HJB}
-i \hbar m \partial_\tau \phi(\tau, \mathbf x)  = \left( i \hbar \gamma^\nu \partial_\nu -  e \gamma^\nu  A_\nu - mc \right)
\left( i \hbar \gamma^\mu \partial_\mu \phi(\tau, \mathbf x) - e \gamma^\mu  A_\mu \phi(\tau, \mathbf x) \right).
\end{equation}

Although this factored equation elegantly generated a Dirac equation without explicitly introducing spin-electromagnetic coupling into the Lagrangian, it nonetheless employed a different spinor representation and, crucially, did not preserve the mass-shell condition. In the standard Dirac equation, acting with the conjugate operator $\gamma^\nu (i \hbar \partial_\nu - e A_\nu) + m c$ on both sides transforms it into a Klein-Gordon-type equation for each spinor component, thereby enforcing the relativistic energy-momentum relation. In the earlier approach, however, the second factor omitted the $+m c$ term. Consequently, its solutions did not satisfy the Klein-Gordon-type equation.

In the present work, by contrast, the standard relativistic square-root dependence on $u^\mu$ is kept in its original nonlinear form and is not linearized. By applying the relativistic “weak condition” $u^\mu u_\mu = c^2$ to the kinetic term of the Lagrangian, rather than linearizing it, we retain the essential $m^2 c^2$ term. The only linearization used later is the Hopf-Cole transform of the resulting nonlinear scalar HJB equation.

This approach aligns with Papiez~\cite{Papiez1981}, who used a similar method to derive the Klein-Gordon equation for a free particle. 
We extend Papiez's idea by incorporating charge-electromagnetic interaction as well as spin-electromagnetic interaction into the Lagrangian. This extension enables the derivation of the Dirac equation while naturally integrating spin dynamics into SOC theory.

The aim of this paper is to derive the Dirac equation from the SOC principles formulated in~\cite{Yordanov2024}, using a specified covariant single-particle Lagrangian as input to the SOC problem while preserving the relativistic mass-shell relation, so that the \(\hbar\to0\) limit of the HJB equation recovers the proper-time relativistic Hamilton--Jacobi equation, whose stationary form yields the classical mass-shell condition, whereas the stationary HJB equation yields the quantum-corrected mass-shell relation.

To illustrate the theory, we consider the Dirac-Landau problem for an electron in a uniform magnetic field as a numerical example. Its role is to demonstrate the analytic optimal drift and the corresponding stochastic action in a standard relativistic solution, rather than to form part of the derivation of the Dirac equation.

In this work we use the following Minkowski space-time metric tensor: $\eta=\text{diag}(+1, -1, -1, -1)$.

\section{Complex Stochastic Optimal Control Theory of Quantum Mechanics}
\label{sec:CSOC_QM}
Historically, SOC originates with Fürth’s~\cite{Furth1933} observation that quantum fluctuations resemble Brownian motion, and Fényes~\cite{Fenyes1952}, who developed a diffusion-based stochastic interpretation recovering the Schrödinger equation. Nelson~\cite{Nelson1966} formalized stochastic mechanics, deriving Schrödinger dynamics from Newtonian mechanics via forward and backward Wiener processes. Yasue~\cite{Yasue1981} and Guerra and Morato~\cite{Guerra1983} cast the theory as stochastic optimal control via variational principles. Papiez~\cite{Papiez1981,Papiez1982} extended it to relativistic particles (yielding Klein-Gordon-type equations for spin‑0) and proposed an early Complex SOC formulation~\cite{Papiez1982}. Pavon~\cite{Pavon1995} introduced a complex quantum velocity while keeping real space-time coordinates and formulated a quantum Hamilton’s principle~\cite{Pavon1995}.

Building on our earlier work~\cite{Yordanov2024b}, we emphasize that allowing the particle to move in four-dimensional complex space-time -- using complex rather than real coordinates -- is essential for mathematical consistency with quantum mechanics.  Conceptually, this amounts to describing the dynamics with two perfectly correlated or anti-correlated Wiener processes, respectively driving the real and imaginary coordinate components, in analogy with Nelson’s forward- and backward-in-time construction~\cite{Nelson1966}.  In their analysis of complex mechanics, Yang \emph{et al.}~\cite{Yang2021_Cian_Dong} showed that restricting the complex velocity to the real axis gives Nelson’s current velocity from the real part, and (up to a sign) the osmotic velocity from the imaginary part.

Consider a particle whose complex four-position $z^{\mu}$ undergoes Brownian motion described by:
\begin{equation}
\begin{aligned}
\label{eq:StochasticProcess}
dz_\mu = w_\mu\, ds + \sigma_\mu\, dW_\mu, \qquad \mu=0,\dots,3,
\end{aligned}
\end{equation}
where $dW_\mu$ denotes the complex Wiener increment, and $\sigma_\mu$ is a constant complex diffusion coefficient proportional to $\sqrt{\hbar/m}$~\cite{Yordanov2024b}.

Within Complex SOC we define the pathwise cost functional (stochastic action) as the time integral of the running cost (single-particle Lagrangian) $L(s,\mathbf z(s),\mathbf w(s))$ on the interval $[\tau_i,\tau_f]$ under control $\mathbf w$ (drift/four-velocity), with $\mathbf z(s)$ the state (complex four-position):
\begin{equation}
\label{eq:def_action}
A\left(\mathbf z_i, \mathbf w(\tau_i \rightarrow \tau_f)\right)
= \int_{\tau_i}^{\tau_f}
L\left(s, \mathbf z(s), \mathbf w(s) \right)\, ds.
\end{equation}
The evolution parameter $s$ in Eq.~\eqref{eq:def_action} need not be the proper time; any strictly increasing parametrization of the worldline is admissible~\cite[Ch.~7.10]{Goldstein2002}. 

The optimal control $\mathbf w^\star(\tau_i\!\rightarrow\!\tau_f)$ is chosen to simultaneously minimize the expectations of the real and imaginary parts of the stochastic action~\cite{Yordanov2024b}. Accordingly, we define the optimal cost over $[\tau_i,\tau_f]$ (minimal expected action):
\begin{equation}
\label{eq:def_min_expected_action}
S(\tau_i,\mathbf z_i;\tau_f) = \min_{\mathbf w(\tau_i \rightarrow \tau_f)}
\Big\langle
  A\left(\mathbf z_i, \mathbf w(\tau_i \rightarrow \tau_f)\right)
\Big\rangle_{\mathbf z_i}.
\end{equation}
Here the expectation is taken over stochastic trajectories with $\mathbf z(\tau_i)=\mathbf z_i$.

For $\tau\in [\tau_i,\tau_f]$, we introduce the value function (the intermediate action):
\begin{equation}
\label{eq:cost_to_go_function}
J(\tau,\mathbf z_\tau; \tau_f) \;=\; \min_{\mathbf w(\tau\to\tau_f)}
\Big\langle \int_{\tau}^{\tau_f} L(s,\mathbf z_s,\mathbf w_s)\, ds \Big\rangle_{\mathbf z_\tau}.
\end{equation}
The value function $J$ satisfies the HJB equation:
\begin{equation}
\label{eq:HJB}
-\partial_\tau J(\tau, \mathbf z; \tau_f)
= \min_{\mathbf w} \left[
L(\tau,\mathbf z,\mathbf w) + w^{\mu}\partial_{\mu}J(\tau,\mathbf z; \tau_f)
+ \frac{1}{2}\sum_{\mu=0}^{3}\sigma_{\mu}\sigma^{\mu}\,\partial_{\mu} \partial_{\mu}J(\tau,\mathbf z; \tau_f)
\right].
\end{equation}
Here the second-order term is the It\^o diffusion term, arising from quadratic variation via It\^o's lemma.

For classical derivations of Eq.~\eqref{eq:HJB}, see the tutorial introduction~\cite{Kappen2011} and the comprehensive monograph~\cite{Fleming2006}.
The complex-valued formulation used here follows~\cite{Yordanov2024b}. 
Note that, within stochastic optimal control, there are two standard formulations: dynamic programming, leading to the stochastic HJB equation~\cite{Bellman1964}, and Pontryagin's Maximum Principle (PMP)~\cite{Pontryagin1987}. 
In this work we follow the HJB formulation.

Once $J$ is obtained as the solution of \eqref{eq:HJB} with terminal condition $J(\tau_f,\mathbf z; \tau_f)=0$, the minimal expected action over $[\tau_i,\tau_f]$ coincides with the value function at the start. 
Substituting $\tau_i$ for $\tau$ in Eq.~\eqref{eq:cost_to_go_function} gives:
\begin{equation}
S(\tau_i, \mathbf z_i; \tau_f) = J(\tau_i, \mathbf z_{\tau_i}; \tau_f).	
\end{equation}

The optimal control at $(\tau, \mathbf z)$ depends on the gradient $\partial_{\mu}J(\tau,\mathbf z; \tau_f)$:
\begin{equation}
\mathbf w^\star(\tau, \mathbf z; \tau_f) \;=\; \arg\min_{\mathbf w} \left(
L(\tau,\mathbf z,\mathbf w) + w^{\mu}\partial_{\mu}J(\tau, \mathbf z; \tau_f)
\right).
\end{equation}	

Complex SOC may be viewed as a stochastic dynamic-programming extension of classical variational mechanics. The optimized quantity is the expected stochastic action of the particle, rather than the classical action associated with a single deterministic path. Within this formulation, the Bellman value function satisfies the HJB equation, while the classical Hamilton--Jacobi equation emerges only in the zero-diffusion, equivalently \(\hbar \to 0\), limit discussed later.

In the next section, we specify the covariant single-particle Lagrangian that serves as the running cost in Eq.~\eqref{eq:HJB} and hence as the defining input to the SOC problem.


\section{Covariant Relativistic Lagrangian}
\label{sec:covariant_lagrangian}

To derive the Dirac equation, we specify a covariant Lagrangian for a relativistic charged particle with spin-$\tfrac12$ in an external electromagnetic field. Within the SOC/HJB framework, this Lagrangian is taken as the input running cost in the scalar HJB equation~\eqref{eq:HJB}.

For the present problem, we take the following Lorentz-covariant Lagrangian for a single electron ($\epsilon = 1$) or positron ($\epsilon = -1$) with mass $m$ and charge $e$ in an electromagnetic field with four-potential $A_\mu$:
\begin{equation}
\label{eq:sqrt_lagrangian}
\Lambda= - m c \sqrt{w_\mu w^\mu} - \epsilon e A_\mu w^\mu - \frac{e\hbar}{2 m} \sigma^{\mu \nu} F_{\mu \nu},
\end{equation}
where $F_{\mu \nu}$ is the electromagnetic field-strength tensor and $\sigma^{\mu \nu}$ is the Dirac spin tensor. We denote the Dirac spin tensor by $\sigma^{\mu\nu}$ (double index). By contrast, $\sigma^\mu$ (single index), introduced in Section~\ref{sec:CSOC_QM}, denotes the diffusion coefficients.

The first two terms in Eq.~\eqref{eq:sqrt_lagrangian} are the standard terms in the covariant least-action formulation of a relativistic charged point particle: the relativistic square-root kinetic term and the minimal electromagnetic coupling.
The third term is absent from the ordinary spinless point-particle action. Rather, covariant spin-field couplings of this general kind appear in action-based formulations of relativistic spinning particles, for example in the Berezin--Marinov framework~\cite{Berezin1977}.
Once inserted into the scalar HJB equation, these terms enter the same value-function dynamics: the kinetic and minimal-coupling terms determine the control-dependent part of the HJB minimization, whereas the Pauli term enters as a state-dependent contribution. 
Although it does not contribute to the first-order stationarity condition because it is independent of the control, it remains in the same HJB equation after substitution of the optimal control and contributes to the later quantum mass-shell relation.

Since $\sigma^{\mu\nu}F_{\mu\nu}(\mathbf z)$ is a $4\times4$ matrix-valued field, the Lagrangian in Eq.~\eqref{eq:sqrt_lagrangian} is not a scalar and cannot be used directly in the scalar HJB equation. To obtain a scalar Lagrangian compatible with the Complex SOC formulation, we introduce a nonzero constant spinor $\chi\in\mathbb{C}^4$, with $\chi^\dagger S\chi\neq0$, and an invertible spin metric $S$ (e.g.\ $S=\gamma^0$). We will refer to $\chi$ as the \emph{scalarization spinor}. Using $\chi$, we define the $\chi$-bilinear scalarization:
\begin{equation}
P_\chi[M]:=\frac{\chi^\dagger S\,M\,\chi}{\chi^\dagger S\,\chi}.
\end{equation}

We then use the scalarized Lagrangian:
\begin{equation}
\label{eq:Lagrangian}
L_\chi(\mathbf z, \mathbf w)=-mc\,\sqrt{w_\mu w^\mu}\;-\;\epsilon e\,A_\mu(\mathbf z)\,w^\mu\;-\;\frac{e\hbar}{2m}\,P_\chi \big[\sigma^{\mu\nu}F_{\mu\nu}(\mathbf z)\big],
\end{equation}
which preserves Lorentz covariance and $U(1)$ gauge invariance.

Only the minimal-coupling term $-\epsilon e\,A_\mu(\mathbf z)\,w^\mu$ enters the HJB minimization, because it depends on the control $\mathbf w$. By contrast, the scalarized Pauli term $P_\chi[\sigma^{\mu\nu}F_{\mu\nu}]$ depends only on the state $\mathbf z$ and not on the control $\mathbf w$, therefore does not modify the pointwise minimization in the HJB equation.

The spinor \(\chi\) is fixed and non-dynamical. It is introduced to produce a scalar Lagrangian compatible with the HJB formulation. In the derivation below it is chosen field-adapted, namely as a spacetime-independent eigenspinor of the spin-field matrix; it then labels the corresponding fixed spin sector. 
The value function \(J_\chi\) appearing in the scalar HJB equation is therefore a scalar Bellman cost-to-go for a fixed scalarization choice \(\chi\), not a dynamical spin variable.

The spin-electromagnetic interaction term does not depend on the sign of the particle's charge in the same way as the minimal-coupling term. As will be shown later, this difference arises because the Pauli term is independent of the particle's velocity and therefore does not contribute to the particle's optimal velocity, whereas the optimal velocity does depend on the sign of the charge and thus distinguishes between electron and positron dynamics.

The spin-electromagnetic interaction term can be written in terms of the Dirac gamma matrices as:
\begin{equation} 
\label{eq:spin_interaction_alt} 
\frac{e \hbar}{2 m} \sigma^{\mu \nu} F_{\mu \nu} = \frac{i e \hbar}{4 m} [\gamma^\mu, \gamma^\nu] F_{\mu \nu},
\end{equation}
where the Dirac spin matrices $\sigma^{\mu \nu}$ are defined in terms of the $\gamma^\mu$ matrices by $\sigma^{\mu \nu} = \frac{i}{2}[\gamma^\mu,\gamma^\nu]$.

The electromagnetic field strength tensor appearing in both Eqs.~\eqref{eq:Lagrangian} and~\eqref{eq:spin_interaction_alt} can be expressed in terms of the electromagnetic potential as follows:
\begin{equation} 
\label{eq:Fmunu} 
F_{\mu \nu} = \partial_\mu A_\nu - \partial_\nu A_\mu.
\end{equation}
 
The components of the four-velocity of the particle are related to the speed of light by the equation:
\begin{equation}
\label{eq:weak_equation}
w_\mu w^\mu = c^2.
\end{equation}
This relation, referred to as the ``weak equation'' by Dirac, allows us to treat $w^\mu$ as unconstrained quantities until all differentiation operations have been carried out, at which point we impose the condition of equation \eqref{eq:weak_equation} (see~\cite{Goldstein2002} Chapter 7.10). This will be the approach we employ as we seek to minimize the expected value of the stochastic action.

\section{Derivation of Dirac Equation}
\label{sec:derivation_of_dirac_equation}
Let us find the minimum of the expected value of the stochastic action using the Lagrangian~\eqref{eq:Lagrangian}. Let us substitute this Lagrangian in~\eqref{eq:HJB}:

\begin{equation}
\label{eq:HJB_sqrt_L}
-\partial_\tau J_\chi =\min_w \left( - mc \sqrt{w_\mu w^\mu} + w^\mu \left(\partial_\mu J_\chi - \epsilon e A_\mu\right) - \frac{e\hbar}{2 m} \mathcal{P}_{\chi}[\sigma^{\mu \nu} F_{\mu \nu}] + \frac{1}{2} \sum_{\mu=0}^3 \sigma^\mu \sigma^\mu \partial_{\mu \mu} J_\chi \right),
\end{equation}
where the subscript on \(J\) indicates that \(J_\chi\) is the value function for the SOC problem defined by the \(\chi\)-bilinear scalarization of the Lagrangian (see Section~\ref{sec:covariant_lagrangian}).

Here \(J_\chi\) is the scalar Bellman value function associated with a fixed scalarization spinor \(\chi\), not a dynamical spin variable. The minimization in Eq.~\eqref{eq:HJB_sqrt_L} is carried out with respect to the control \(w^\mu\) only. Consequently, the Pauli term \(-\tfrac{e\hbar}{2m}\mathcal P_{\chi}[\sigma^{\mu \nu} F_{\mu \nu}]\) enters the HJB equation as a state-dependent scalar term, but because it is independent of \(w^\mu\), it does not contribute to the first-order stationarity condition for the minimizer.

Differentiating only the \(w\)-dependent terms with respect to $w^\mu$, we obtain the optimal control $w^\mu$ at which the expression attains a minimum:
\begin{equation}
- m c \frac{2 w^{\star,{(\chi)}}_\mu}{2 \sqrt{w^{\star,{(\chi)}}_\mu w^\mu_{\star,{(\chi)}}}} + \left( \partial_\mu J_\chi - \epsilon e A_\mu \right) = 0.
\end{equation}

If we use the weak condition~\eqref{eq:weak_equation} then we derive:

\begin{equation}
 - w^{\star, (\chi)}_\mu + \frac{1}{m} \left( \partial_\mu J_\chi - \epsilon e A_\mu\right)= 0.
\end{equation}

For brevity, we denote the optimal control by $w_\mu$ and omit the superscript $w^\star_\mu$ throughout:

\begin{equation}
\label{eq:optimal_control_J}
w_\mu^{(\chi)} =  \frac{1}{m} \left( \partial_\mu J_\chi - \epsilon e A_\mu \right).
\end{equation}

It is then natural to identify the gauge-invariant kinetic four-momentum as:
\begin{equation}
  \pi_\mu^{(\chi)} \;\equiv\; \partial_\mu J_\chi \;-\; \epsilon e\, A_\mu .
  \label{eq:pLink}
\end{equation}
The superscript \((\chi)\) on the optimal control \(w_\mu^{(\chi)}\) and on the kinetic four-momentum \(\pi_\mu^{(\chi)}\) indicates dependence on the scalarization spinor \(\chi\) via the \(\chi\)-bilinear scalarization of the Lagrangian.

For brevity, in the derivation below we omit the \(\chi\) labels. All quantities are understood to be evaluated for a fixed scalarization spinor \(\chi\).

Under a local \(U(1)\) gauge transformation with gauge function \(\xi(\mathbf z)\),
\(A_\mu \to A_\mu + \partial_\mu \xi(\mathbf z)\) and \(J \to J + \epsilon e\,\xi(\mathbf z)\),
so \(\mathbf \pi\) is gauge invariant by construction: \(\pi_\mu \to \pi_\mu\).

Once we have performed differentiation operations to find the optimal control policy~\eqref{eq:optimal_control_J} we can apply the weak equation~\eqref{eq:weak_equation} into Eq.~\eqref{eq:HJB}. Since the four-velocity is optimal, we can then remove the minimization function from~\eqref{eq:HJB_sqrt_L} to obtain:
\begin{equation}
\label{eq:HJB_linear-vec}
-\partial_\tau J =  - mc^2 + w^\mu \left(\partial_\mu J - \epsilon e A_\mu \right) - \frac{e\hbar}{2 m} \mathcal{P}_{\chi}[ \sigma^{\mu \nu} F_{\mu \nu}] + \frac{1}{2} \sum_{\mu = 0}^3 \sigma^\mu \sigma^\mu \partial_\mu \partial_\mu J.
\end{equation}

We note that a similar equation, without electromagnetic coupling and spin terms, was developed by Papiez~\cite{Papiez1981}.
However, he used it to derive the Klein-Gordon equation for a free particle in the absence of electromagnetic fields.

In our previous work on Complex SOC theory in quantum mechanics~\cite{Yordanov2024b} we obtained the following componentwise relation for the complex diffusion:
\begin{equation}
\label{eq:diffusion_coef_sqr}
\sigma^\mu \sigma^\mu  =2 i \epsilon \eta^{\mu \mu}  \frac{\hbar}{m}, \quad \mu=0,1,2,3.
\end{equation}
The sign $\epsilon$ in the equation aligns with the sign of the charge in the Lagrangian~\eqref{eq:sqrt_lagrangian}. It indicates that perfect correlation $\epsilon = 1$ or perfect anti-correlation $\epsilon = -1$ between the real and imaginary stochastic processes correspond, respectively, to particles or antiparticles. In other words, switching from perfect correlation to anti-correlation effectively changes the sign of the charge.

Note that in Nelson mechanics~\cite{Nelson1966} the diffusion coefficient for both the forward and backward stochastic processes is $\sigma_{\mathrm N}=\sqrt{\hbar/(2m)}$. 
Yang \emph{et al.}~\cite{Yang2021_Cian_Dong}, in their complex mechanics, likewise use the Nelson value $\sigma_{\mathrm N}$ for the real and imaginary stochastic components. 
By contrast, the Complex SOC theory developed in~\cite{Yordanov2024b} and used here employs the diffusion coefficient \(\sigma=\sqrt{\hbar/m}\).

If we substitute equations~\eqref{eq:diffusion_coef_sqr} and \eqref{eq:optimal_control_J} into~\eqref{eq:HJB_linear-vec} and multiply by $m$ both sides of the equation we obtain:

\begin{equation}
\begin{aligned}
\label{eq:HJB_J}
- m \partial_\tau J = &- m^2 c^2 +i \epsilon \hbar \partial^\mu \partial_\mu J - \frac{e\hbar}{2 } \mathcal P_\chi [\sigma^{\mu \nu} F_{\mu \nu}]  + \left(\partial^\mu J - \epsilon e A^\mu \right)^2.
\end{aligned}
\end{equation}

Let us introduce a new function $\tilde{J}$:

\begin{equation}
\label{eq:S_definition}
J= i \epsilon \hbar \tilde{J}.
\end{equation}

Substituting Eq.~\eqref{eq:S_definition} into Eq.~\eqref{eq:optimal_control_J}, we obtain the following expression for the optimal control velocity:
\begin{equation}
\label{eq:optimal_control_u_vec}
w_\mu = \frac{\epsilon}{m} \left(i  \hbar \partial_\mu \tilde{J} - e A_\mu \right).
\end{equation}
This result explicitly shows that the optimal velocity depends on the sign of the particle's charge, $\epsilon$.

Substituting Eq.~\eqref{eq:S_definition} into Eq.~\eqref{eq:HJB_J} we obtain:
\begin{equation}
\begin{aligned}
\label{eq:HJB_tilda_J}
- i \epsilon \hbar m \partial_\tau \tilde J = &- m^2 c^2 - \hbar^2 \partial^\mu \partial_\mu \tilde J - \frac{e\hbar}{2} \mathcal P_\chi [ \sigma^{\mu \nu} F_{\mu \nu} ] + \left(i \hbar \partial^\mu \tilde  J - e A^\mu \right)^2.
\end{aligned}
\end{equation}

Note that in the last equation the charge sign (or, equivalently, the correlation/anti-correlation coefficient $\epsilon$) cancels out due to the squaring operation. This motivates our choice not to include the charge sign in the spin-electromagnetic term of the Lagrangian defined in Section~\ref{sec:covariant_lagrangian}.

Expanding the last term of the above equation we obtain:
\begin{equation}
\begin{aligned}
- i \epsilon \hbar m \partial_\tau \tilde{J} = &- m^2 c^2 - \hbar^2\left( \partial^\mu \tilde{J}\partial_\mu \tilde{J} + \partial^\mu \partial_\mu \tilde{J} \right) - \frac{e\hbar}{2} \mathcal{P}_{\chi}[\sigma^{\mu \nu} F_{\mu \nu}] - 2 i \hbar \partial^\mu \tilde{J} e A_\mu + e^2 A^\mu A_\mu.
\end{aligned}
\end{equation}

Introduce the Hopf-Cole map $\tilde J=\ln\phi$. Using
$\partial^\mu\tilde J\,\partial_\mu\tilde J+\partial^\mu\partial_\mu\tilde J=(\partial^\mu\partial_\mu\phi)/\phi$,
Eq.~\eqref{eq:HJB_tilda_J} becomes linear in $\phi$:
\begin{equation}
\begin{aligned}
\label{eq:HJB_divided_by_phi}
-i \epsilon\hbar m \frac{\partial_\tau \phi}{\phi} =& - m^2 c^2 - \hbar^2 \frac{\partial^\mu \partial_\mu \phi}{\phi} - \frac{e\hbar}{2} \mathcal{P}_{\chi}[\sigma^{\mu \nu} F_{\mu \nu}] - 2 e A^\mu i \hbar \frac{\partial_\mu \phi}{\phi} + e^2 A^\mu A_\mu.
\end{aligned}
\end{equation}

Multiplying both sides of Eq.~\eqref{eq:HJB_divided_by_phi} by $\phi$ and noting that $\phi$ is a scalar (so it commutes with the Dirac-matrix factors), using Eq.~\eqref{eq:spin_interaction_alt}, finally we obtain:

\begin{equation}
\begin{aligned}
\label{eq:hjb_dirac_initial}
-i \epsilon \hbar m \partial_\tau \phi = & - m^2 c^2 \phi - \hbar^2 \partial^\mu \partial_\mu \phi - \frac{i e \hbar}{4} \mathcal{P}_{\chi}[[\gamma^\mu, \gamma^\nu] F_{\mu \nu} \phi] - 2 i e \hbar A^\mu \partial_\mu \phi + e^2 A^\mu A_\mu \phi.
\end{aligned}
\end{equation}

We will show that equation \eqref{eq:hjb_dirac_initial} is equivalent to:

\begin{equation}
\label{eq:hjb_dirac_final}
-i \epsilon \hbar m \partial_\tau \phi =\mathcal{P}_{\chi}[ \left( i \hbar \gamma^\nu \partial_\nu -  e \gamma^\nu  A_\nu - mc \right)
\left(i \hbar \gamma^\mu \partial_\mu  - e \gamma^\mu  A_\mu  + m c \, \right) \phi].
\end{equation}

Expanding the right side before applying \(P_\chi\) gives:

\begin{equation}
\begin{aligned}
\label{eq:all_terms}
&- m^2 c^2 \phi - \hbar^2 \gamma^\nu \gamma^\mu \partial_\nu \partial_\mu \phi - \\
& - i  e \hbar \gamma^\nu \gamma^\mu \left( \partial_\nu A_\mu \right)  \phi - i e \hbar \gamma^\nu \gamma^\mu  A_\mu \partial_\nu \phi - \\
& - i e \hbar \gamma^\nu \gamma^\mu A_\nu \partial_\mu \phi + e^2 \gamma^\nu \gamma^\mu  A_\nu A_\mu \phi + \\
& + mc \left( i \hbar \gamma^\mu \partial_\mu \phi - e \gamma^\mu  A_\mu \phi \right) - \\
& - mc \left( i \hbar \gamma^\nu \partial_\nu \phi - e \gamma^\nu  A_\nu \phi \right).
\end{aligned}
\end{equation}

To simplify the above equation we will use the following relations:
\begin{equation}
	\gamma^\nu \gamma^\mu = \frac{1}{2} \{ \gamma^\nu, \gamma^\mu \} + \frac{1}{2} [ \gamma^\nu, \gamma^\mu ] = \eta^{\nu \mu} \mathbb{I} + \frac{1}{2} [ \gamma^\nu, \gamma^\mu ].
\end{equation}

Since $\partial_\mu \partial_\nu$ is a symmetric tensor, the second term simplifies to:
\begin{equation}
\begin{aligned}
\label{eq:term_2}
-\hbar^2 \gamma^\nu \gamma^\mu \partial_\nu \partial_\mu  \phi = -\hbar^2  \partial^\mu \partial_\mu  \phi.
\end{aligned}
\end{equation}

In a similar way, the sixth term simplifies to:
\begin{equation}
\begin{aligned}
\label{eq:term_6}
e^2 \gamma^\nu \gamma^\mu A_\nu A_\mu  \phi = e^2 A^\mu A_\mu  \phi.
\end{aligned}
\end{equation}

The third term is:
\begin{equation}
\begin{aligned}
\label{eq:term_3}
-i e \hbar \gamma^\nu \gamma^\mu  (\partial_\nu A_\mu )\phi &= -i e \hbar \gamma^\nu \gamma^\mu  \left(\frac{1}{2}(\partial_\nu A_\mu  + \partial_\mu A_\nu) + \frac{1}{2}(\partial_\nu A_\mu  - \partial_\mu A_\nu) \right) \phi \\
& = - i e \hbar \frac{1}{2} (\partial_\mu A^\mu)  \phi - \frac{1}{4} [\gamma^\mu, \gamma^\nu] (\partial_\mu A_\nu - \partial_\nu A_\mu)  \phi\\
& = - \frac{i e \hbar}{4} [\gamma^\mu, \gamma^\nu] F_{\mu \nu} \phi,
\end{aligned}
\end{equation}
where we have used the Lorenz gauge condition~\cite{Lorenz1867} $\partial_\mu A^\mu=0$.

The forth and fifth terms are:
\begin{equation}
\begin{aligned}
\label{eq:term_4_5}
-i e \hbar \gamma^\nu \gamma^\mu (A_\mu \partial_\nu + A_\nu \partial_\mu) \phi = -2 i e \hbar A^\mu \partial_\mu  \phi.
\end{aligned}
\end{equation}

The seventh and eighth terms cancel each other out.

If we substitute Eqs.~\eqref{eq:term_2}, ~\eqref{eq:term_3}, ~\eqref{eq:term_4_5} into Eq.~\eqref{eq:all_terms} it will turn into equation~\eqref{eq:hjb_dirac_initial}, which proves that both equations \eqref{eq:hjb_dirac_initial} and \eqref{eq:hjb_dirac_final} are equivalent.

\subsection{Squared-Dirac equation in a fixed spin sector}
\label{sec:squared-dirac}
Keeping the scalarization spinor $\chi$ constant (see Section~\ref{sec:covariant_lagrangian}), derivatives act only on the scalar \(\phi\) and on the fixed external gauge field \(A_\mu\).
We now choose \(\chi\) to be field-adapted, namely a nonzero constant spinor satisfying
\begin{equation}
\label{eq:field_adapted_chi}
\sigma^{\mu\nu}F_{\mu\nu}(\mathbf z)\,\chi=f_\chi(\mathbf z)\,\chi
\end{equation}
throughout the spacetime region considered. For the uniform magnetic field used below, one may take \(\chi=v_s\), where \(\Sigma_z v_s=s\,v_s\) with \(s=\pm1\) (Appendix~\ref{app:landau}).

For a scalar \(\phi\), the expansion in Eqs.~\eqref{eq:all_terms}--\eqref{eq:term_4_5} consists of terms proportional to the identity in spinor space together with the Pauli term. By Eq.~\eqref{eq:field_adapted_chi}, the Pauli term also maps \(\chi\) into the same one-dimensional spin sector. Hence the squared operator preserves the sector spanned by \(\chi\). Since the scalarization in Eq.~\eqref{eq:hjb_dirac_final} yields the corresponding eigenvalue in this sector, the scalar HJB equation lifts to the non-stationary squared-Dirac equation
\begin{equation}\label{eq:unprojected_squared}
-\,i\,\varepsilon\,\hbar m\,\partial_\tau(\phi\,\chi) = \big(i\hbar\gamma^\nu\partial_\nu-e\gamma^\nu A_\nu-mc\big)
\big(i\hbar\gamma^\mu\partial_\mu-e\gamma^\mu A_\mu+mc\big)\,(\phi\,\chi).
\end{equation}
The present scalar construction is therefore restricted to backgrounds for which a spacetime-independent eigenspinor satisfying Eq.~\eqref{eq:field_adapted_chi} exists.

\subsection{Dirac Equation}

With the field-adapted \(\chi\) of Section~\ref{sec:squared-dirac}, define the spinor:
\begin{equation}
\label{eq:psi_tau}
\psi(\tau, \mathbf z) = \big( i \hbar \gamma^\mu \partial_\mu \phi(\tau, \mathbf z)  - e \gamma^\mu  A_\mu \phi(\tau, \mathbf z) + mc \, \phi(\tau, \mathbf z) \big)\chi.
\end{equation}

Then, using the non-stationary squared Dirac equation, Eq.~\eqref{eq:unprojected_squared}, we obtain:

\begin{equation}
\label{eq:HJB_Dirac}
-i \epsilon \hbar m \, \partial_\tau \big(\phi(\tau, \mathbf z)\,\chi\big)= \left( i \hbar \gamma^\nu \partial_\nu - e \gamma^\nu  A_\nu - mc \right) \psi(\tau, \mathbf z).
\end{equation}

Note that equation~\eqref{eq:unprojected_squared} is a second-order differential equation. Fock~\cite[Chap. 37-2]{Fock2019} arrived at this same equation, as well as the spinor in Eq.~\eqref{eq:psi_tau}, while investigating proper time in the Dirac equation. In this paper, we derive both, starting from the fundamental principles of Complex SOC theory.

The Dirac equation is obtained in the stationary case by setting the proper time derivative to zero.

\begin{equation}
\label{eq:Dirac_equation}
\begin{aligned}
& \left( i \hbar \gamma^\nu \partial_\nu - e \gamma^\nu  A_\nu - mc \right) \psi(\mathbf z) = 0.
\end{aligned}
\end{equation}

Here, $\bm{\psi}(\mathbf z)$ denotes the stationary solution, which does not depend on proper time.
As can be seen from~\eqref{eq:Dirac_equation} that $\bm{\psi}(\mathbf z)$ is the Dirac spinor, which from~\eqref{eq:psi_tau} becomes:

\begin{equation}
\label{eq:psi_phi}
\psi(\mathbf z) = \big( i \hbar \gamma^\mu \partial_\mu \phi(\mathbf z) - e \gamma^\mu  A_\mu \phi(\mathbf z) + mc \, \phi(\mathbf z) \big)\chi.
\end{equation}

Introduce the first-order Dirac operator:
\begin{equation}
D_{+} \;:=\; i\hbar\,\gamma^\mu \partial_\mu \;-\; e\,\gamma^\mu A_\mu \;+\; mc .
\end{equation}
Then the Dirac spinor can be written compactly as:
\begin{equation}
\psi(\mathbf z) \;=\; D_{+}\,\big(\phi(\mathbf z)\,\chi\big).
\end{equation}

\subsection{Summary of the derivation}
To summarize the path of the long derivation, we display the main flow of equations leading to the Dirac equation, and the flow by which the Dirac spinor \(\psi\) and the kinetic four-momentum \( \mathbf \pi \) are introduced. For brevity, \(\chi\)-labels are suppressed. 
All statements hold for a fixed field-adapted scalarization spinor \(\chi\), satisfying Eq.~\eqref{eq:field_adapted_chi}.

\paragraph*{Equations flow.}
\[
\begin{array}{c}\text{It\^o SDE }\\ \text{ for } \mathbf z(\tau) \end{array}
\xrightarrow{\ \text{SOC, DP} }
\begin{array}{c}\text{HJB }\\ \text{ for } J(\tau, \mathbf z) \end{array}
\xrightarrow{\ \text{Hopf-Cole }\ }
\begin{array}{c}\text{squared-Dirac}\\ \text{equation for }\phi(\tau, \mathbf z)\,\chi\end{array}
\xrightarrow{\ \psi:=D_{+}(\phi\chi)\ }
\begin{array}{c}\text{Dirac equation}\\ \text{ for }\psi(\mathbf z)\end{array}
\]

The particle's motion is governed by a four-dimensional It\^o SDE for the spacetime coordinate \(\mathbf z\in\mathbb{C}^4\) with drift four-velocity \(\mathbf w\). 
In the SOC view, \(\mathbf z\) is the state and \(\mathbf w\) is the admissible control.
We formulate a SOC problem with running cost given by the single-particle Lagrangian \(L(\tau, \mathbf z,\mathbf w)\).
Dynamic programming (DP) and It\^o's lemma yield the HJB for \(J(\tau,\mathbf z)\)~\cite{Kappen2011}.
Because the Pauli spin-field term is matrix-valued, we employ the \(\chi\)-bilinear scalarization (Section~\ref{sec:covariant_lagrangian}) to keep the HJB equation scalar.

For a field-adapted \(\chi\), the Hopf-Cole transform yields the non-stationary squared-Dirac equation for \(\phi(\tau,\mathbf z)\,\chi\) (Section~\ref{sec:squared-dirac}).

We then define the Dirac spinor \(\psi\) by applying the first-order Dirac operator \(D_{+}\) to \(\phi\,\chi\), i.e. \(\psi=D_{+}(\phi\chi)\), and in the proper time stationary case obtain the Dirac equation.

\paragraph*{Dirac spinor and kinetic four-momentum flow.}
\begin{equation*}
\begin{aligned}
\mathbf z(\tau) &\xrightarrow{\ \text{SOC, DP}\ } J(\tau, \mathbf z)
            \xrightarrow{\ \text{Hopf-Cole}\ } \phi(\tau, \mathbf z)\,\chi
            \xrightarrow{\ D_{+}(\cdot)\ } \psi(\mathbf z)
\\[0.6ex]
\mathbf z(\tau) &\xrightarrow{\ \text{SOC, DP}\ } J(\tau, \mathbf z)
            \xrightarrow{\ \arg\min_{\mathbf w} \ }\; \mathbf w^\star(\tau, \mathbf z)
            \xrightarrow{\ m } \mathbf \pi(\tau, \mathbf z)
\end{aligned}
\end{equation*}

The first flow mirrors the equation flow described above.
In parallel, a solution $J$ of the HJB yields the optimal feedback $\mathbf w^\star$
 via the pointwise minimization
The kinetic four-momentum is defined as \(\pi_\mu=m w_\mu = \partial_\mu J - eA_\mu\) and \( \mathbf \pi \) is gauge invariant under \( A_\mu \to A_\mu + \partial_\mu \xi \) with \( J \to J + e\,\xi \).

\section{Bilinear physical observables and the It\^{o} quadratic-covariation correction}
\label{sec:ito-observables}

To connect the SOC dynamics to measurable quantities, we define how to evaluate
bilinear observables along It\^o trajectories.

In Complex SOC theory we evolve fully complex coordinates $z_\tau^\mu$ by the It\^{o} SDE in Eq.~\eqref{eq:StochasticProcess}.
When a quantity involves a bilinear, state-dependent factor $F(z)\,G(z)$, the plain pointwise product is biased by quadratic variation.
Using the quadratic covariation (bracket) notation~\cite{RevuzYor1999}, the It\^{o} product rule reads:
\begin{equation}
  d(FG) \;=\; F\,dG \;+\; G\,dF \;+\; d\langle F,G\rangle_\tau,
\end{equation}
and for Eq.~\eqref{eq:StochasticProcess} with constant, diagonal diffusion:
\begin{equation}
  d\langle F(z),G(z)\rangle_\tau
  \;=\; \sum_{\mu}(\sigma^\mu)^2\,\partial_\mu F\!\big(z_\tau\big)\,\partial_\mu G\!\big(z_\tau\big)\,d\tau.
  \label{eq:ito-bracket-constsigma}
\end{equation}

We therefore define the physical bilinear locally by the It\^{o} quadratic-covariation correction:
\begin{equation}
  (FG)^{\mathrm{phys}}(z)
  \;:=\;
  F(z)\,G(z)
  \;+\;\frac{1}{2}\sum_{\mu}(\sigma^\mu)^2\,\partial_\mu F(z)\,\partial_\mu G(z),
  \label{eq:ito-counterterm-def}
\end{equation}
so that along a path the instantaneous observable is $(FG)^{\mathrm{phys}}\!\big(z_\tau\big)$.
We use the superscript “phys” because this time-symmetric correction removes the quadratic-variation bias implied by the bracket, preserves the usual chain and product rules under smooth changes of variables, and reduces to the ordinary product as the diffusion tends to zero.

This prescription is used  in the next Section~\ref{sec:quantum_mass_shell} to interpret the quantum mass-shell relation and, in Section~\ref{sec:numerics} and Appendix~\ref{app:landau-ito-half}, to evaluate the It\^o covariation correction to the electromagnetic term \(L_{\mathrm{EM}}=e\,A_\mu w^\mu\).

\section{Classical and quantum mass shell}
\label{sec:classical_mass_shell}

In this section we develop the proper-time Hamilton--Jacobi form of the classical and quantum mass-shell relations. The classical relation is obtained from the \(\hbar\to0\) limit of the HJB equation, while the stationary HJB equation yields the corresponding quantum-corrected mass-shell relation.

Throughout this section, for brevity we drop the $\chi$ labels from $J_\chi$, $\pi_{(\chi)}^\mu$, and $\phi_\chi$, writing $J$, $\pi^\mu$, and $\phi$. All identities are understood in one fixed field-adapted spin sector, with $\chi$ satisfying Eq.~\eqref{eq:field_adapted_chi}.
\subsection{Classical mass shell (HJ limit)}
With $\pi_\mu = \partial_\mu J - \epsilon e A_\mu$ (see Eq.~\eqref{eq:pLink}),
the $\hbar\!\to\!0$ limit of Eq.~\eqref{eq:HJB_J} is the proper time Hamilton-Jacobi (HJ) equation:
\begin{equation}
\label{eq:HJ_limit}
-\,m\,\partial_\tau J \;=\; -\,m^2 c^2 \;+\; \pi_\mu \pi^\mu.
\end{equation}
In the stationary case $\partial_\tau J=0$, Eq.~\eqref{eq:HJ_limit} reduces to the classical mass shell:
\begin{equation}
\label{eq:classical_mass_shell}
\pi_\mu \pi^\mu \;=\; m^2 c^2 .
\end{equation}

\subsection{Quantum mass shell (stationary HJB)}
\label{sec:quantum_mass_shell}
In the stochastic-quantum setting developed here, starting from the nonlinear HJB equation~\eqref{eq:HJB_tilda_J} and imposing stationarity in proper time ($\partial_\tau \tilde J = 0$) -- equivalently, the stationary squared-Dirac equation -- yields the quantum-corrected mass-shell relation:
\begin{equation}
\label{eq:sqr_dirac}
\left(i\hbar\,\partial_\mu \tilde J - e A_\mu\right)
\!\left(i\hbar\,\partial^\mu \tilde J - e A^\mu\right)
\;-\;\hbar^2\,\partial^\mu\partial_\mu \tilde J	
\;-\;\frac{e\hbar}{2}\,\mathcal P_\chi [\sigma^{\mu\nu}F_{\mu\nu}]
\;=\; m^2 c^2 .
\end{equation}
Under the Hopf-Cole identification $\tilde J=\ln\phi$ and Eq.~\eqref{eq:pLink} the above reduces to:
\begin{equation}
\label{eq:quantum_mass_shell_phi}
\pi_\mu \pi^\mu
\;=\; m^2 c^2 \;+\;\hbar^2\,\partial^\mu\partial_\mu \ln\phi
\;+\;\frac{e\hbar}{2}\, \mathcal P_\chi [\sigma^{\mu\nu}F_{\mu\nu}].
\end{equation}

The Pauli term in Eq.~\eqref{eq:quantum_mass_shell_phi} is inherited directly from the spin-field contribution
$-\frac{e\hbar}{2m}\,P_\chi[\sigma^{\mu\nu}F_{\mu\nu}(z)]$ in the scalarized Lagrangian of Section~\ref{sec:covariant_lagrangian}.
Because it is independent of the control/drift, it contributes to the scalar quantum mass-shell only as an additive invariant correction.

The classical mass shell Eq.~\eqref{eq:classical_mass_shell} is recovered from Eq.~\eqref{eq:quantum_mass_shell_phi} when the It\^o diffusion correction, also called the covariant quantum potential~\cite{Nikolic2005}, and the Pauli term vanish, as in the WKB/semiclassical limit~\cite{Papiez1981}. In general, these quantum corrections are essential for exact on-shell consistency.

\subsection{Identification with It\^o covariation of \(\pi^2\)}
\label{sec:ito_cov}
We now connect Eq.~\eqref{eq:quantum_mass_shell_phi} to the local notion of physical bilinears in Section~\ref{sec:ito-observables}. Applying Eq.~\eqref{eq:ito-counterterm-def} with \(F=\pi_\mu\), \(G=\pi^\mu\) gives the pointwise definition:
\begin{equation}
\label{eq:pi2-phys-local}
\big(\pi^\mu \pi_\mu\big)^{\mathrm{phys}}(z)
\;:=\;
\pi^\mu(z)\,\pi_\mu(z)
\;+\;
\frac{1}{2}\sum_{\nu}(\sigma^\nu)^2\,
\partial_\nu \pi_\mu(z)\,\partial_\nu \pi^\mu(z),
\end{equation}
with \((\sigma^\nu)^2\) from Eq.~\eqref{eq:diffusion_coef_sqr} and
\(\pi_\mu=i\epsilon\hbar\,\partial_\mu\ln\phi-\epsilon e\,A_\mu\).
Averaging along the trajectory yields the corresponding mean observable:
\begin{equation}
\label{eq:phys-pi2-def}
\big\langle \pi^\mu \pi_\mu \big\rangle^{\mathrm{phys}}
\;=\;
\big\langle \pi^\mu \pi_\mu \big\rangle
\;+\;
\frac{1}{2}\sum_{\nu}(\sigma^\nu)^2
\Big\langle \partial_\nu \pi_\mu\,\partial_\nu \pi^\mu \Big\rangle .
\end{equation}

Expanding Eq.~\eqref{eq:phys-pi2-def}, applying the chain rule, and splitting
\(\partial_\nu A_\mu\) into symmetric and antisymmetric parts exactly as in the
derivation from Eq.~\eqref{eq:all_terms} to Eq.~\eqref{eq:term_3} one finds: the antisymmetric piece yields \(F_{\mu\nu}\),
while the symmetric piece is a total divergence. 
Under averaging with vanishing boundary flux, the divergence term vanishes by
integration by parts, and the remaining contribution cancels the average of the
quantum potential. Therefore:
\begin{equation}
\label{eq:chi_mass_shell_avg}
\big\langle \pi^\mu \pi_\mu \big\rangle^{\mathrm{phys}}
= m^2 c^2
+\Big\langle \frac{e\hbar}{2}\,\mathcal P_\chi [\sigma^{\mu\nu}F_{\mu\nu}]\Big\rangle .	
\end{equation}

Thus, the quantum correction in the mass-shell relation is precisely the It\^o covariation contribution for \(\pi\!\cdot\!\pi\): the covariant quantum-potential term is canceled by the covariation, while the antisymmetric part yields the Pauli term. 
To our knowledge, this is the first explicit identification -- in a Lorentz-covariant, spinful setting -- equating the Dirac mass-shell quantum correction with the It\^o covariation of the bilinear \(\pi\!\cdot\!\pi\) observable.

\subsection{Charge-flip and branch exchange on the mass shell}
We write the physical charge as $q=\epsilon e$, and the same $\epsilon=\pm 1$ appears in the Lagrangian, Eq.~\eqref{eq:Lagrangian}.
As recalled in Eq.~\eqref{eq:diffusion_coef_sqr}, this $\epsilon$ also enters the diffusion coefficients and encodes the correlation between the real and imaginary parts of the complex stochastic process: $\epsilon=+1$ (perfect correlation) and $\epsilon=-1$ (perfect anticorrelation)~\cite{Yordanov2024b}.

Recall the $\chi$-bilinear scalarized quantum mass shell from Eq.~\eqref{eq:quantum_mass_shell_phi} and define:
\begin{equation}
\label{eq:S_scalar_def}
S(\mathbf z)\;:=\;m^2 c^2
\;+\;\hbar^2\,\partial^\mu\partial_\mu \ln\phi(\mathbf z)
\;+\;\frac{e\hbar}{2}\,\mathcal P_\chi\!\big[\sigma^{\mu\nu}F_{\mu\nu}(\mathbf z)\big],
\end{equation}
so that the mass shell can be written as $\pi^2(\epsilon;\mathbf z)=S(\mathbf z)$. Here $\pi_\mu(\epsilon;\mathbf z)$ is the gauge-invariant kinetic four-momentum from Eq.~\eqref{eq:pLink}. Using Eq.~\eqref{eq:S_definition}, one has:
\begin{equation}
\label{eq:pi_momentum}
\pi_\mu(\epsilon;\mathbf z)=\epsilon\left(i\hbar\,\partial_\mu\tilde J(\mathbf z)-eA_\mu(\mathbf z)\right).
\end{equation}
It follows that the kinetic four-momentum is odd in the charge sign:
\begin{equation}
\label{eq:pi_sign_symmetry}
\pi_\mu(-\epsilon;\mathbf z)=-\,\pi_\mu(\epsilon;\mathbf z),
\end{equation}
while $S(\mathbf z)$ carries no dependence on $\epsilon$.

Introduce the matrix fields:
\begin{equation}
\label{eq:Mtilde_def}
M_\pm(\epsilon;\mathbf z)\;:=\;\gamma^\mu \pi_\mu(\epsilon;\mathbf z)\ \pm\ \sqrt{S(\mathbf z)},
\end{equation}
where $\sqrt{S}$ denotes any fixed local branch of the scalar square root. 
Henceforth, for brevity, we suppress the explicit $\mathbf z$-dependence of $\pi$, $S$, and $M$.

By the Clifford identity $\{\gamma^\mu,\gamma^\nu\}=2\eta^{\mu\nu} \mathbb I_4$, we have: $\big(\gamma^\mu \pi_\mu(\epsilon)\big)^2=\pi^\mu(\epsilon)\,\pi_\mu(\epsilon)\,\mathbb I_4$, hence:
\begin{equation}
\label{eq:algebraic_fact}
\big(\gamma^\mu \pi_\mu(\epsilon)-\sqrt{S}\big)\,
\big(\gamma^\mu \pi_\mu(\epsilon)+\sqrt{S}\big)
\;=\;\big(\pi^2(\epsilon)-S\big)\,\mathbb I_4.
\end{equation}

The mass-shell condition \eqref{eq:quantum_mass_shell_phi} then implies:
\begin{equation}
\label{eq:product_zero}
M_-(\epsilon)\,M_+(\epsilon)\;=\;0.
\end{equation}

Consequently, for any $\mathbf z$ with $\pi^2(\epsilon;\mathbf z)=S(\mathbf z)$, at least one of $M_\pm(\epsilon)$ has a nontrivial kernel.
We may thus attach algebraic branches by choosing nonzero spinors $\psi_\pm$ such that:
\begin{equation}
\label{eq:algebraic_branches}
\begin{aligned}
&M_+(\epsilon)\,\psi_+=0
\quad\text{(``$+s$'' algebraic branch)},\\
&M_-(\epsilon)\,\psi_-=0
\quad\text{(``$-s$'' algebraic branch)}.
\end{aligned}
\end{equation}

Equivalently, letting $s:=\sqrt{S}$ (any local branch), the branch spinors are pointwise eigen-spinors of $\gamma^\mu\pi_\mu(\epsilon)$ with eigenvalues $\pm s$:
\begin{equation}
\gamma^\mu\pi_\mu(\epsilon)\,\psi_\pm=\pm s\,\psi_\pm.
\end{equation}
We therefore call the eigenspaces of $\gamma^\mu\pi_\mu(\epsilon)$ corresponding to the eigenvalues $+s$ and $-s$ the $\pm s$ branches.

In the absence of external electromagnetic fields and in the stationary case, the additional terms in $S$ vanish; hence $S=m^2c^2$ and $s=mc$, and we will refer to the branches as the $\pm mc$ branches.
With $\pi_\mu$ constant, this reproduces the standard momentum-space Dirac equations
$(\gamma^\mu p_\mu \pm mc)\,u/v=0$ from the Dirac operator formalism.
These two branches coincide with the usual positive and negative energy sheets $E = \pm mc^2$ of the Dirac mass shell. 
In the interacting Complex SOC theory, $S(z)$ is complex and $\pm s(z)$ are complexified branches that reduce to these sheets when the quantum and Pauli corrections vanish.

Using Eq.~\eqref{eq:pi_sign_symmetry} and noting that $S$ is independent of $\epsilon$ we find:
\begin{equation}
\label{eq:M_swap}
 M_\pm(-\epsilon)
\;=\;-\big(\gamma^\mu \pi_\mu(\epsilon)\ \mp\ \sqrt{S}\big)
\;=\;-\, M_{\mp}(\epsilon).
\end{equation}

Under a charge flip the kernel condition swaps,
$M_+(\epsilon)\,\psi=0 \;\Longleftrightarrow\; M_-(-\epsilon)\,\psi=0$, the mass-shell relation Eq.~\eqref{eq:quantum_mass_shell_phi} remains invariant, and the branches are exchanged.

\section{Non-stationary HJB with free proper time boundary conditions}
\label{sec:nonstationary-HJB-choice-of-w}

Starting from the non-stationary HJB equation~\eqref{eq:hjb_dirac_final}, a convenient separation ansatz is:
\begin{equation}
\label{eq:phi-tau-separation}
\phi(\tau,\mathbf z)\;=\;\exp\!\Big(-\,\frac{i\,\lambda}{\hbar m}\,\tau\Big)\,\tilde\phi(\mathbf z),
\end{equation}
where \(\lambda\) is a separation constant fixed by proper time boundary conditions.
The spatial factor $\tilde\phi$ satisfies the squared-Dirac eigenproblem:
\begin{equation}
\label{eq:sq-Dirac-eig}
\Big(i \hbar \gamma^\nu \partial_\nu -  e \gamma^\nu  A_\nu - mc\Big)
\Big(i \hbar \gamma^\mu \partial_\mu  - e \gamma^\mu  A_\mu  + mc\Big)\,(\tilde\phi\,\chi)
= \lambda\,(\tilde\phi\,\chi),
\end{equation}
so, in general, $\tilde\phi$ depends on $\lambda$. The case $\lambda=0$ corresponds to the homogeneous squared-Dirac equation.

In the Complex SOC formulation the optimal drift is defined from the non-stationary scalar \(\phi(\tau,\mathbf z)\) via the Hopf-Cole map \(J=i\epsilon\hbar\ln\phi\) and Eq.~\eqref{eq:optimal_control_u_vec}:
\begin{equation}
\label{eq:w-from-phi-tau}
w_\mu(\tau,\mathbf z)\;=\;\frac{\epsilon}{m}\Big(i\hbar\,\partial_\mu\ln\phi(\tau,\mathbf z)\;-\;e\,A_\mu(\mathbf z)\Big).
\end{equation}
Using Eq.~\eqref{eq:phi-tau-separation} in Eq.~\eqref{eq:w-from-phi-tau} yields:
\begin{equation}
\label{eq:w-components-with-lambda}
w_\mu(\mathbf z)\;=\;\frac{\epsilon}{m}\Big(i\hbar\,\partial_\mu\ln\tilde\phi(\mathbf z)\;-\;e\,A_\mu(\mathbf z)\Big).
\end{equation}
hence $w^\mu$ is, in general, a $\lambda$-dependent object (through $\tilde\phi$) unless the proper time boundary conditions fixing $\lambda$ are specified.

When $\tilde\phi$ is a stationary first-order Dirac eigenstate with
eigen-energy $E(\lambda)$, its $z^0$-phase is fixed, hence:
\begin{equation}
\label{eq:w0-from-energy}
w^0(\mathbf z)\;=\;\frac{\epsilon}{m}\Big(\frac{E(\lambda)}{c}\;-\;e\,A^0(\mathbf z)\Big).
\end{equation}
In this work we set $\lambda = 0$ and write $E \equiv E(0)$. With this choice
$E$ coincides with the real energy parameter appearing in the stationary
squared Dirac equation in the given electromagnetic background, and
Eq.~\eqref{eq:w0-from-energy} provides the temporal drift component $w^0(\mathbf z)$.

\section{Numerical demonstration: electron in a uniform magnetic field}
\label{sec:numerics}

In this section we present numerical experiments for the Complex SOC formulation of relativistic quantum mechanics developed in Sections~\ref{sec:CSOC_QM}--\ref{sec:derivation_of_dirac_equation}, using the Dirac-Landau problem -- an electron in a uniform magnetic field -- only as an illustrative benchmark. It is not part of the logical derivation of the Dirac equation or of the mass-shell relations. We compare the average kinetic, electromagnetic, spin, and total action components computed under the optimal drift \(w^\star\) with the analytic Dirac-Landau values summarized in Appendix~\ref{app:LandauSOC}. We then examine local minimality, evaluating whether the average stochastic action for trajectories with a common initial space-time point is minimized at the analytically derived optimal drift \(w^\star_\mu\).

In both numerical experiments, reconstruction of wavefunctions or spectra is not attempted. 
That task amounts to a high-dimensional optimization over drift fields (or control parameters) and is computationally demanding on classical hardware. Looking ahead, variational quantum algorithms -- such as the Variational Quantum Eigensolver (VQE)~\cite{Peruzzo2014} and the Quantum Approximate Optimization Algorithm (QAOA)~\cite{Farhi2014} -- may help locate minima of a trajectory-ensemble-averaged, action-defined cost functional, estimated from circuit samples, even in more complex electromagnetic field configurations.

\subsection{Physical model, optimal control, and stochastic action}
Starting from the general Lagrangian in Section~\ref{sec:CSOC_QM}, we specialize to a uniform magnetic field along the \(z\)-axis and adopt the Landau gauge:
\begin{equation}
A^\mu=(0,0,Bx,0).
\end{equation}

Substituting this potential into Eq.~\eqref{eq:sqrt_lagrangian} yields the split Lagrangian \(L=L_{\rm kin}+L_{\rm EM}+L_{\rm sp}\):
\begin{equation}
\label{eq:Lsplit}
L_{\rm kin}=-mc\,\sqrt{w^\mu w_\mu}+mc^2,\qquad
L_{\rm EM}=-e\,A_\mu w^\mu=e\,B\,x\,w^{y},\qquad
L_{\rm sp}=-(g/4)\,\hbar\omega_c\,s,
\end{equation}
where the terms represent, respectively, the kinetic energy, charge-field interaction, and spin-field coupling; here \(\omega_c=|e|B/m\), \(s=\pm1\) is the spin projection, and \(w^{y}\) is the \(y\)-component of the spatial drift.

For the field-adapted scalarization, we choose $\chi=v_s$, where $\Sigma_zv_s=s\,v_s$ with $s=\pm1$. The spin-field term in Eq.~\eqref{eq:Lsplit} is then evaluated in the corresponding fixed spin sector (see Appendix~\ref{app:landau}).

For a given quantum state \((n,s)\), the optimal drift \(w^{\mu\,\star}\) follows from Eq.~\eqref{eq:optimal_control_J} together with the Hopf-Cole transform \(J = i\hbar \ln \phi_{n,s}\), where \(\phi_{n,s}\) denotes the scalar factor of the Landau spinor listed in Appendix~\ref{app:LandauSOC}:
\begin{equation}
\label{eq:w_k_star}
  w^{\mu\,\star}(\mathbf z)
    = \frac{i\hbar}{m}\,\partial^\mu \!\ln\phi_{n,s}(\mathbf z)
    - \frac{e}{m}A^\mu(\mathbf z), \quad \mu=0,\cdots,3.
\end{equation}

The stochastic action accumulated along a trajectory over a time horizon \(T\) is:
\begin{equation}
\label{eq:stochastic_action}
  S = \int_0^T \Big( L_{\rm kin}(w^\mu) + L_{\rm EM}(z^\mu,w^\mu) + L_{\rm sp}(s) \Big)\,dt .
\end{equation}
This quantity is the primary observable used in the numerical analysis below.

\subsection{Numerical method}
\label{subsec:numerical_method}

We perform two complementary numerical experiments:

\begin{itemize}
\item \textbf{Optimal action values} (Section~\ref{subsec:optimal_actions}).
Compute the average kinetic, electromagnetic, spin, and total action components under the optimal drift \(w^\star\) and compare them with the analytic Dirac-Landau values (Appendix~\ref{app:LandauSOC}) for sufficiently long horizons \(T>T_c\).

\item \textbf{Local optimality of the action} (Section~\ref{subsec:local_opt}).
Assess whether the average stochastic action is minimized at the optimal drift \(w^\star_\mu\) by perturbing admissible controls within the mass-shell tangent space.
\end{itemize}

All numerical results reported here were generated with the publicly available codebase~\cite{Yordanov2025Code}.

\paragraph{Euler-Maruyama evolution.}
We discretize time on a uniform grid \(t_k=k\Delta t\) with \(k=0,\cdots,N_t\) and \(\Delta t=T/N_t\).
The four-position \(z^\mu\) evolves under the controlled It\^o SDE and is advanced by an Euler-Maruyama step:
\begin{equation}
\label{eq:euler_maruyama}
  z_{k+1} = z_k + w_k\,\Delta t
  + \sigma\,(I - i\,\eta)\,\xi_k\,\sqrt{\Delta t},
  \qquad
  \sigma=\sqrt{\hbar/m},\quad \xi_k\sim\mathcal N(\mathbf 0, I_4),
\end{equation}
with \(w_k = w^\star(z_k)\) for the optimal run, or \(w_k = w^\star(z_k)+\delta_k\) for perturbed runs.

\paragraph{Stochastic action.}
The discrete stochastic action is obtained from the Riemann sum approximation of Eq.~\eqref{eq:stochastic_action}:
\begin{equation}
\label{eq:S_discrete}
  S \;\approx\;
  \sum_{k=0}^{N_t-1}
    \big( L_{\rm kin}(w_k) + L_{\rm EM}(z_k,w_k) + L_{\rm sp}(s) \big)\,\Delta t .
\end{equation}
For comparison with the Dirac-Landau spectrum we also report the dimensionless normalization:
\begin{equation}
\label{eq:action_norm}
  S_{\mathrm{norm}} \equiv \frac{S}{\hbar \omega_c T}.
\end{equation}
In Section~\ref{subsec:local_opt} we use the unnormalized \(S\) to study local optimality, whereas in Section~\ref{subsec:optimal_actions} we report \(S_{\mathrm{norm}}\).

\paragraph{Perturbed controls.}
We probe local optimality around the optimal control $w^\star$ via small perturbations $w_k=w^\star_k+\delta_k$.
In the Complex SOC/HJB derivation the control $w^\mu$ is unconstrained during optimization and the weak relation $w^\mu w_\mu=c^2$ is imposed only after minimization.
For the small-deviation test we restrict admissible directions by first-order feasibility of the weak relation:

\emph{Admissible first-order tests.}
Let $g(w)=w^\mu w_\mu-c^2$ and $w(\varepsilon)=w^\star+\varepsilon\delta+O(\varepsilon^2)$.
Enforcing $g(w(\varepsilon))\equiv0$ to first order gives
$\frac{d}{d\varepsilon} g\big(w(\varepsilon)\big)\Big|_{\varepsilon=0}=2\,w^{\star} \cdot \delta =0$, i.e. $\delta$ lies in the mass-shell tangent space at $w^\star$.

\emph{Phase-fixed test path.}
Since $w^\star\!\cdot\!\delta=0$ is homogeneous, tangent directions are defined only up to a global phase: $(\varepsilon,\delta)\sim(\varepsilon e^{i\theta},e^{-i\theta}\delta)$ yield the same $\Delta w=\varepsilon\delta$.
We fix this phase freedom by choosing the representative with $q\equiv\delta \cdot \delta \in\mathbb R_{<0}$ and take $\varepsilon\in\mathbb R$.
Concretely, we set the phase so that $\arg q=\pi$ and then normalise $\delta\leftarrow\delta/\sqrt{|q|}$ so that $\delta \cdot \delta \approx-1$; the scalar $\varepsilon$ then solely sets the perturbation amplitude.
This phase convention does not restrict the set of admissible perturbations; it fixes a redundant parameterization and provides a frame-independent one-parameter slice for the local test.

\emph{Relativistic feasibility.}
With mostly negative metric signature, causality requires a timelike four-drift: $\Re(w^\mu w_\mu)>0$.
Using $w^{\star2}=c^2$ and $w^\star\!\cdot\!\delta=0$, along $w(\varepsilon)=w^\star+\varepsilon\delta$ one has:
\[
(w^\star+\varepsilon\delta)^2 \;=\; c^2+\varepsilon^2 q,\qquad q=\delta \cdot \delta,
\]
so timelikeness amounts to $c^2+\varepsilon^2\Re(q)>0$.
Under the above phase convention $q\in\mathbb R_{<0}$, hence the radicand is positive real for sufficiently small $|\varepsilon|$, and the principal square-root update is well posed and future-directed.

\paragraph{Control distance.}
We quantify deviations from the optimal drift using the Minkowski distance:
\begin{equation}
\label{eq:Rdelta_unnorm}
  R_{\delta}
  = \frac{1}{N_t}\sum_{k=0}^{N_t-1}
    \sqrt{\bigl|(w_k-w^\star_k)^{\!\top}\eta\,(w_k-w^\star_k)\bigr|},
\end{equation}
and its dimensionless form:
\begin{equation}
\label{eq:Rdelta_norm}
  R_{\delta,\mathrm{norm}} \equiv \frac{R_{\delta}}{c}.
\end{equation}

\subsection{Simulation of optimal action values}
\label{subsec:optimal_actions}

Using the Complex SOC theory developed above, we perform stochastic simulations under the optimal drift \(w^\star\) and test agreement with the analytic Dirac-Landau values within statistical uncertainty.
The numerical algorithm is as follows.

\paragraph{Numerical algorithm.}
\begin{enumerate}
  \item \textbf{Inputs.}
  Select the state \((n,s)\), magnetic field \(B\), time horizon \(T\), number of steps \(N_t\) (so \(\Delta t=T/N_t\)), and number of trajectories \(N_{\rm traj}\).
  All runs use horizons \(T>T_c\) and keep \(T\) below the stability/convergence limit discussed in the numerical stability Section~\ref{subsec:numerical_stability}.

  \item \textbf{Initialization.}
  Initialize the four-position as:
  \begin{equation}
  	  z_0=(0,\,X_0+\ell_B,\,0,\,0),
  \end{equation}
  where \(X_0=-\hbar k_y/(eB)\) is the guiding-center coordinate.
  Offsetting by one magnetic length \(\ell_B=\sqrt{\hbar/(|e|B)}\) avoids numerical instabilities near Landau wavefunction nodes.

  \item \textbf{Propagation under the optimal drift.}
  For each trajectory and each step \(k=0,\cdots,N_t-1\), evaluate \(w^\star_\mu(n,s;z_k)\) using Eq.~\eqref{eq:w_k_star}; and advance the state with the Euler-Maruyama update (see Eq.~\eqref{eq:euler_maruyama}).

  \item \textbf{Action accumulation and normalization.}
  Accumulate the discrete action components using Eq.~\eqref{eq:S_discrete}, then apply the normalization in Eq.~\eqref{eq:action_norm} to obtain the reported quantities \(S_{\mathrm{kin}}, S_{\mathrm{EM}}, S_{\mathrm{sp}}, S_{\mathrm{tot}}\).

  \item \textbf{Ensemble averages and uncertainty (95\% CI).}
  For each action component (kinetic, electromagnetic, spin, total), compute the mean value for many trajectory runs:
  \(\langle S_{\mathrm{kin}}\rangle\), \(\langle S_{\mathrm{EM}}\rangle\), \(\langle S_{\mathrm{sp}}\rangle\), and \(\langle S_{\mathrm{tot}}\rangle\).
  Construct \(95\%\) confidence intervals via a nonparametric bootstrap over trajectories:
  draw \(N_{\mathrm{boot}}\) resamples (with replacement) of size \(N_{\rm traj}\), compute the corresponding resampled means, and take the percentile interval \([\mathrm{q}_{2.5},\,\mathrm{q}_{97.5}]\).
  
      \item \textbf{Physical action computation.}
  Evaluate observables as in Section~\ref{sec:ito-observables}.
  For the electromagnetic piece this means adding the It\^o covariation counterterm.
  In the Dirac-Landau case ($k_y=0$) this gives a universal $+\tfrac12$ shift (see Appendix~\ref{app:landau-ito-half}): \(\langle S_{\rm EM}/\hbar\omega_c T\rangle^{\rm phys}=\langle S_{\rm EM}/\hbar\omega_c T\rangle +\tfrac12.\)

The kinetic running cost (Eq.~\eqref{eq:Lsplit}) is nonlinear in $w$ and, after inserting $w=w^\star(z)$, becomes a nonlinear state-dependent observable sampled along It\^{o} trajectories.
We therefore evaluate $S_{\rm kin}$ using the It\^{o} nonlinear square-root prescription of Appendix~\ref{app:kinetic-closure-eigenstate}, which closes the kinetic sector on shell and yields $S_{\rm kin}^{\rm phys}=0$ within numerical accuracy.

The spin term $L_{\mathrm{sp}}$ is constant for fixed $s$ and uniform $B$.
    
    \item \textbf{Analytic comparison.}
  Compare each mean action with the analytic Dirac-Landau value listed in Appendix~\ref{app:LandauSOC}; declare agreement when that value lies within the bootstrap \(95\%\) confidence interval.
\end{enumerate}

\subsection{Simulation of local optimality}
\label{subsec:local_opt}

Building on the baseline simulation in Section~\ref{subsec:optimal_actions}, we probe local minimality by perturbing the optimal control within the mass-shell tangent space and measuring the excess action relative to \(w^\star\).
The algorithm for the local-optimality test is as follows.

\paragraph{Numerical algorithm.}
\begin{enumerate}
  \item \textbf{Setup and initial condition.}
  Fix \(B\), \((n,s)\), \(T\), and \(N_t\); set the uniform time grid \(t_k=k\Delta t\) with \(\Delta t=T/N_t\). 
  Choose the initial point \(z_0\) as specified above (outside nodal regions).

  \item \textbf{Optimal control and baseline action.}
  At each step, evaluate the optimal control \(w^\star\) via Eqs.~\eqref{eq:w_k_star}; propagate with the Euler-Maruyama update (Eq.~\eqref{eq:euler_maruyama}) and accumulate the action (Eq.~\eqref{eq:S_discrete}).
  Generate \(N_{\mathrm{opt}}\) independent trajectories under the unperturbed \(w^\star\) and compute the baseline mean \(\langle S^\star \rangle\) as in Section~\ref{subsec:optimal_actions}.

  \item \textbf{Admissible perturbations in the mass-shell tangent space.}
  Replace the analytic drift by \(w_k = w^\star_k + \delta_k\), where \(\delta_k\) are admissible SOC-consistent deviations (see Section~\ref{subsec:numerical_method}).
  Operationally, at each time step start from a deterministic seed \(\delta_{\rm raw}\) and project to the tangent space of the mass shell:
  \begin{equation}
  \label{eq:minkowski_projection}
    \delta_k \;=\; \delta_{\rm raw}
    \;-\; w^\star_k\, \frac{\,w^{\star\mu}_k\,\eta_{\mu\nu}\,\delta_{\rm raw}^{\nu}\,}{\,w^{\star\alpha}_k\,\eta_{\alpha\beta}\,w^{\star\beta}_k\,}\,,
  \end{equation}
  reseeding if this projection is numerically degenerate.
    The distance \(R_{\delta,\mathrm{norm}}\) of perturbated control from the optimal one is computed using Eq.~\eqref{eq:Rdelta_unnorm}.
  \item \textbf{Action averaging.}
  For each perturbed control, average over \(N_{\mathrm{avg}}\) independent trajectories to compute \(\langle S\rangle\), define:
  \[
      \Re\langle \Delta S \rangle \;=\; \Re\langle S\rangle \;-\; \Re \langle S^\star \rangle.
  \]
  \item \textbf{Reporting.}
  Plot \(\Re\langle \Delta S \rangle\) versus \(R_{\delta,\mathrm{norm}}\) to visualize the rise in action away from the optimum.
\end{enumerate}

\subsection{Numerical results: optimal action values}

We computed average stochastic action components for selected Dirac-Landau states \((n,s)\) at \(B=0.1\,\mathrm{T}\), using \(N_{\rm traj}=10000\) trajectories and a total of \(N_t=10000\) time steps over a horizon \(T=5\,T_c\), where \(T_c=2\pi/\omega_c\).
This choice followed a preliminary stability scan (see Section~\ref{subsec:numerical_stability}), which showed that averaging over several cyclotron periods is sufficient to suppress residual transients and discretization effects.

\begin{table}[ht]
\centering
\small
\caption{Physical action components, normalized by \(\hbar\omega_c T\), computed by the Complex SOC simulation at \(B=0.1\,\mathrm{T}\); \(N_{\rm traj}=10000\); \(N_t=10000\) total steps; \(T=5\,T_c\). Analytic Dirac-Landau values are shown in the last column.}
\label{tab:optimal_actions}
\begin{tabular}{c c c c c}
\toprule
State $(n,s)$ & Component & {Mean} & {95\% CI} & {Theory} \\
\midrule
\multirow{4}{*}{$(0,-1)$} 
  & $\langle S_{\rm EM}\rangle^{\rm phys}$  & -0.4826 & [-0.5100, -0.4545] & -0.5000 \\
  & $\langle S_{\rm SP}\rangle^{\rm phys}$  & \phantom{-}0.5000 & [\phantom{-}0.5000, \phantom{-}0.5000] & \phantom{-}0.5000 \\
  & $\langle S_{\rm kin}\rangle^{\rm phys}$ & \phantom{-}0.0000 & [\phantom{-}0.0000, \phantom{-}0.0000] & \phantom{-}0.0000 \\
  & $\langle S_{\rm tot}\rangle^{\rm phys}$ &  \phantom{-}0.0174 & [-0.0100, \phantom{-}0.0455] & \phantom{-}0.0000 \\
\midrule
\multirow{4}{*}{$(0,+1)$} 
  & $\langle S_{\rm EM}\rangle^{\rm phys}$  & -0.4829 & [-0.5113, -0.4532] & -0.5000 \\
  & $\langle S_{\rm SP}\rangle^{\rm phys}$  & -0.5000 & [-0.5000, -0.5000] & -0.5000 \\
  & $\langle S_{\rm kin}\rangle^{\rm phys}$ & \phantom{-}0.0000 & [\phantom{-}0.0000, \phantom{-}0.0000] & \phantom{-}0.0000 \\
  & $\langle S_{\rm tot}\rangle^{\rm phys}$ & -0.9829 & [-1.0113, -0.9532] & -1.0000 \\
\midrule
\multirow{4}{*}{$(1,-1)$} 
  & $\langle S_{\rm EM}\rangle^{\rm phys}$  & -1.4927 & [-1.5220, -1.4646] & -1.5000 \\
  & $\langle S_{\rm SP}\rangle^{\rm phys}$  & \phantom{-}0.5000 & [\phantom{-}0.5000, \phantom{-}0.5000] & \phantom{-}0.5000 \\
  & $\langle S_{\rm kin}\rangle^{\rm phys}$ &  \phantom{-}0.0000 & [\phantom{-}0.0000, \phantom{-}0.0000] & \phantom{-}0.0000 \\
  & $\langle S_{\rm tot}\rangle^{\rm phys}$ & -0.9927 & [-1.0220, -0.9646] & -1.0000 \\
\midrule
\multirow{4}{*}{$(1,+1)$} 
  & $\langle S_{\rm EM}\rangle^{\rm phys}$  & -1.4922 & [-1.5209, -1.4640] & -1.5000 \\
  & $\langle S_{\rm SP}\rangle^{\rm phys}$  & -0.5000 & [-0.5000, -0.5000] & -0.5000 \\
  & $\langle S_{\rm kin}\rangle^{\rm phys}$ & \phantom{-}0.0000 & [\phantom{-}0.0000, \phantom{-}0.0000] & \phantom{-}0.0000 \\
  & $\langle S_{\rm tot}\rangle^{\rm phys}$ & -1.9922 & [-2.0209, -1.9640] & -2.0000 \\
\bottomrule
\end{tabular}
\end{table}

Across all states \((n,s)\) listed in Table~\ref{tab:optimal_actions}, the spin component is \(\pm 0.5\) exactly, since \(L_{\rm sp}\) is constant (see Eq.~\eqref{eq:Lsplit}).
The kinetic contribution from Eq.~\eqref{eq:Lsplit} requires an explicit prescription because it is a nonlinear square-root observable evaluated on the state-dependent drift $w^\star(z)$.
We therefore adopt the It\^{o} nonlinear square-root prescription of Appendix~\ref{app:kinetic-closure-eigenstate}, which implements the on-shell closure in the kinetic sector and enforces $S_{\rm kin}^{\rm phys}(n,s)=0$.
With this prescription, the reported kinetic contribution is zero within numerical accuracy.
The electromagnetic component exhibits the expected Landau-level dependence, and the resulting totals are consistent with the analytic Dirac-Landau expressions in Appendix~\ref{app:LandauSOC}.

The simulation results show that, at the optimal drift \(w^\star\), the Complex SOC formulation reproduces the Dirac-Landau decomposition of the action within the reported \(95\%\) confidence intervals.


\subsection{Numerical results: local optimality of the stochastic action}
\label{subsec:numerical_results}

\begin{figure}[ht!]
    \centering
    \subfloat{
        \includegraphics[width=0.48\linewidth]{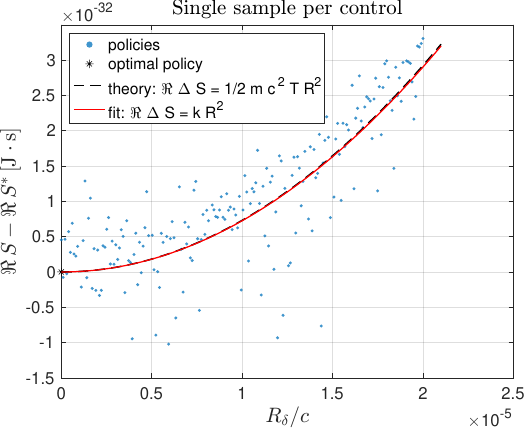}
    }
    \hfill
    \subfloat{
        \includegraphics[width=0.48\linewidth]{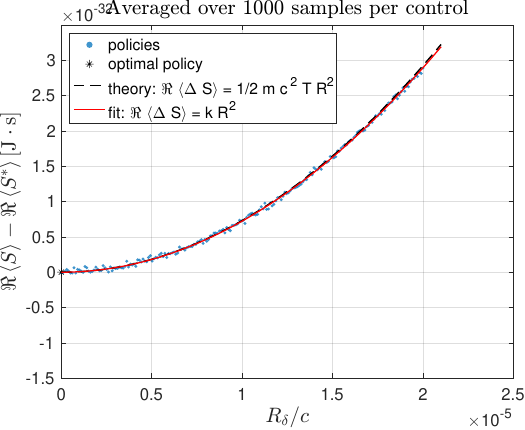}
    }
    \caption{Excess action versus normalized control distance \(R_{\delta}/c\) for perturbations around the optimal control \(w^\star\). Left: single realizations, plotting \(\Re\, \Delta S\), where stochastic fluctuations may yield \(S<S^\star\). Right: averages over \(N_{\mathrm{avg}}=1000\) trajectories per control, plotting \(\Re \,\langle \Delta S\rangle\), exhibiting a quadratic minimum at \(R_{\delta}/c=0\). Parameters: \(B=0.1\,\mathrm{T}\), \(n=0\), \(s=-1\).}
    \label{fig:bowl}
\end{figure}

In this section we present numerical results from simulations that probe the local minimality described in Section~\ref{subsec:local_opt}.
The results are shown in figure~\ref{fig:bowl}, which displays the dependence of the excess expected action \( \Re\langle \Delta S\rangle \)
on the normalized control distance \(R_{\delta,\mathrm{norm}}\).

The left panel shows single realizations: pathwise fluctuations may yield \(S<S^\star\), which is consistent with stochastic variability and does not contradict optimality, defined at the level of the expectation.
Averaging over many trajectories per control, as shown in the right panel with \(N_{\mathrm{avg}}=1000\), suppresses sample-to-sample noise and reveals the expected quadratic dependence, with a minimum at \(R_{\delta}/c=0\).

The curvature is determined by the relativistic kinetic term.
Expanding:
\begin{equation}
	L_{\rm kin} = -mc\,\sqrt{w^\mu w_\mu}+mc^2	
\end{equation}
around \(w^\star_\mu\) for a small spacelike perturbation \(\delta^\mu\) tangent to the mass shell, so that \(w^{\star\mu}\eta_{\mu\nu}\delta^\nu=0\) and \(\Re(\delta^\mu\eta_{\mu\nu}\delta^\nu)<0\), yields, to quadratic order:
\begin{equation}
	\Re \,\langle \Delta S \rangle = \tfrac{1}{2}\, m c^2 T \, R_{\delta,\mathrm{norm}}^{2}.	
\end{equation}

The numerical results confirm the Complex SOC prediction that, for trajectories initiated at a common space-time point, the average stochastic action attains a local minimum at the optimal control \(w^\star\).

\subsection{Numerical stability}
\label{subsec:numerical_stability}

\begin{figure}[ht!]
  \centering
  \includegraphics[width=0.67\linewidth]{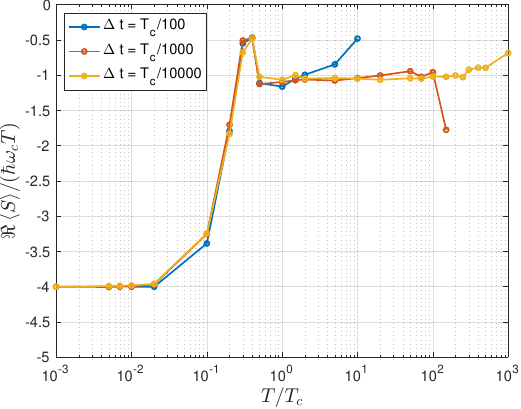}
  \caption{Time-step stability of the normalized stochastic action.
  Mean \(\Re\langle S\rangle/(\hbar\omega_c T)\) versus \(T/T_c\) for three time-step choices:
  \(\Delta t=T_c/100\), \(\Delta t=T_c/1000\), and \(\Delta t=T_c/10000\).
  Finer steps preserve agreement over longer horizons, whereas a coarse step eventually loses accuracy.
  Parameters: \(B=0.1\,\mathrm{T}\), \(n=0\), \(s=+1\), \(N_{\rm run}=1000\) trajectories per point.}
  \label{fig:stability_dt_vs_T}
\end{figure}

To assess time-discretization effects we fixed the drift to the optimal value \(w^\star\) and varied the horizon \(T\).
The time step was tied to the cyclotron period \(T_c\) by fixing the number of steps per period \(N_{T_c}\in\{100,\,1000,\,10000\}\), i.e. \(\Delta t = T_c/N_{T_c}\).
For each \((T,N_{T_c})\) we ran \(N_{\rm run}=1000\) trajectories and reported the mean normalized stochastic action \(\Re\langle S\rangle/(\hbar\omega_c T)\).
The dependence of \(\Re\langle S\rangle/(\hbar\omega_c T)\) on \(T/T_c\) for the three step sizes is presented in figure~\ref{fig:stability_dt_vs_T}.

We observe three distinct regimes:
\begin{itemize}
  \item \emph{Short horizons.}
  The normalized total starts near \(-4\).
  This follows from the initial condition \(x_0=X_0+2\ell_B\) with \(k_y=0\) so \(X_0=0\): at \(t=0\) one has \(L^{\rm phys}_{\rm EM}/(\hbar\omega_c)=-3.5\) and \(L_{\rm sp}/(\hbar\omega_c)=-0.5\) (for \(s=+1\)), while \(L_{\rm kin}=0\) by the It\^{o} nonlinear square-root prescription of Appendix~\ref{app:kinetic-closure-eigenstate}, hence the initial total is \(-4\).
  As \(T\) increases and averaging over the cyclotron motion sets in, the value relaxes toward the Dirac-Landau prediction determined by \((n,s)\).

  \item \emph{Resolved regime.}
  Once the step size \(\Delta t\) resolves the cyclotron dynamics, the mean stabilizes close to the analytic value over a broad range of \(T/T_c\).
  The finer steps \(\Delta t=T_c/1000\) and \(T_c/10000\) produce an extended flat plateau, while the coarser step \(\Delta t=T_c/100\) yields a narrower flat region.

  \item \emph{Loss of accuracy at coarse step.}
  For large \(T/T_c\) with \(\Delta t=T_c/100\), the curve drifts away from the plateau.
  The coarse step under-resolves the periodic product \(x\,w^y\) in the electromagnetic contribution and introduces a per-period quadrature bias that accumulates over many periods, so the Euler-Maruyama scheme no longer tracks the complex drift-diffusion dynamics accurately.
  Using a smaller \(\Delta t\) pushes this breakdown to longer horizons.
\end{itemize}

\section{Conclusion}

In this work, we have derived the Dirac equation from SOC principles using a specified covariant single-particle input Lagrangian, in which the nonlinear relativistic kinetic term is retained in its original form and the Pauli spin-electromagnetic coupling is incorporated through scalarization in a fixed spin sector selected by a spacetime-independent eigenspinor of \(\sigma^{\mu\nu}F_{\mu\nu}\).

Crucially, this approach restores the mass-shell condition that was lost in earlier linearized treatments. The resulting spinor representation naturally matches the Fock representation of the Dirac spinor, as introduced by Fock~\cite{Fock2019} through his proper-time approach. These findings not only resolve previous inconsistencies but also underscore the power of stochastic optimal control in offering a robust bridge between stochastic mechanics and relativistic quantum theory.

Beyond the analytic derivation, we illustrate the theory with stochastic simulations of the Dirac-Landau problem. At the analytic optimal drift \(w^\star\), we find that the average stochastic action attains a local minimum: deterministic perturbations with small control distance confirm local minimality. For horizons exceeding a few cyclotron periods, the average action components under \(w^\star\) are consistent with the Dirac-Landau values within statistical uncertainty.

The Complex SOC theory provides a first-principles formulation of relativistic quantum mechanics within the fixed-sector setting considered here. It casts the recovery of spectra and wavefunctions as a high-dimensional optimization over drift fields. Variational quantum-classical algorithms -- such as VQE and QAOA variants with action-defined cost functionals -- are natural candidates for exploration. Extending the scalar construction to backgrounds without a spacetime-independent eigenspinor is left for future work.

\appendix
\section{Dirac-Landau Problem: Eigenfunctions, Spectrum, and Action Decomposition}
\label{app:LandauSOC}

Consider an electron of mass \(m\) and charge \(e<0\) in a uniform magnetic
field \(\mathbf B=B\,\hat{\mathbf z}\) and Landau gauge \(A^\mu=(A^0,\mathbf A)=(0,0,Bx,0)\).
We use the standard constants: 
\(\omega_c=|e|B/m\) (cyclotron frequency), \(\ell_B=\sqrt{\hbar/(|e|B)}\) (magnetic length),
\(X_0=-\,\hbar k_y/(eB)\) (guiding-center), and \(\xi=(x-X_0)/\ell_B\) (shifted coordinate).

\subsection{Stationary eigenfunctions and spectrum}
\label{app:landau}
The stationary (squared-Dirac) eigenfunctions~\cite{Landau1965,berestetskii1982} factorize as:
\begin{equation}
\label{eq:phi_const_B}
\Phi_{n,k_y,p_z,s}(x,y,z)
= \phi_{n,k_y,p_z,s}(x,y,z)\,v_s,
\qquad
\phi_{n,k_y,p_z,s}(x,y,z)
= \mathcal N_n\,H_n(\xi)\,e^{-\xi^2/2}\;
e^{\,i(k_y y+p_z z)/\hbar},
\qquad
\mathcal N_n=\frac{(|e|B/\pi\hbar)^{1/4}}{\sqrt{2^n n!}},
\end{equation}
where \(n=0,1,2,\dots\), \(H_n\) are Hermite polynomials, and \(v_s\) is a constant
spinor with \(\Sigma_z v_s=s\,v_s\) (\(s=\pm1\)). 

For the fixed-sector scalarization used in this paper we take \(\chi=v_s\). Since $\sigma_{\mu\nu}F^{\mu\nu}=2\,\Sigma_z\,B$, the corresponding scalarization is
\begin{equation}\label{eq:proj_uniformB}
P_{v_s}\!\big[\sigma_{\mu\nu}F^{\mu\nu}\big]=2sB.
\end{equation}

The relativistic Landau spectrum
(including the \(g=2\) spin splitting generated by minimal coupling) is:
\begin{equation}
\label{eq:LandauEnergy_SI}
E_{n,s}=\pm\sqrt{\,m^2c^4+p_z^2c^2+(2n+1+s)\, \hbar\omega_c\,m c^2\,}\,.
\end{equation}

\subsection{Gradients of \texorpdfstring{$\ln\phi$}{ln φ}}
The spatial derivatives needed for the computation of optimal control $w^\star$ are:
\begin{equation}
\label{eq:dxlnphi}
\partial_x\ln\phi
=
\frac{1}{\ell_B}\!\left(\frac{H_n'(\xi)}{H_n(\xi)}-\xi\right)
=
\begin{cases}
-\,(x-X_0)/\ell_B^2, & n=0,\\[6pt]
\dfrac{1}{\,x-X_0\,}-\dfrac{x-X_0}{\ell_B^2}, & n=1,
\end{cases}
\end{equation}
\begin{equation}
\label{eq:dylnphi}
\partial_y\ln\phi= i\,k_y/\hbar,
\qquad
\partial_z\ln\phi= i\,p_z/\hbar.
\end{equation}

\subsection{Action splits from operator expectations}
\label{sec:operator_action_split}
We decompose the one-particle action over a time horizon \(T\) as
\begin{equation}
\label{eq:S_split}
	S_{\rm tot}=S_{\rm kin}+S_{\rm EM}+S_{\rm SP},
\end{equation}
where each piece is the
time-integral of the corresponding operator expectation in a stationary
Dirac eigenstate \(|\Psi_{n,k_y,p_z,s}\rangle\) with energy \(E_{n,s}\).

We set \(\dot{\mathbf x}=\frac{i}{\hbar}[H,\mathbf x]=c\,\boldsymbol{\alpha}\),
\(\boldsymbol{\Pi}=\mathbf p-e\mathbf A\).

\paragraph{Electromagnetic contribution.}
\begin{equation}
\label{eq:Sem_def}
S_{\rm EM}
=\int_0^T\!\big\langle -\,e\,\mathbf A(\hat{\mathbf x})\cdot\dot{\mathbf x}\big\rangle\,dt
=-\,eBc\int_0^T\!\langle \hat{x}\,\alpha_y\rangle\,dt.
\end{equation}
In the Landau eigenstates one has \(\langle \hat{x}\rangle=X_0\) and
\(\langle(\hat{x}-X_0)^2\rangle=(n+\tfrac12)\ell_B^2\); using the Heisenberg
equation \(\dot{\Pi}_x=eB\,\dot y=eBc\,\alpha_y\) and stationarity,
\(\frac{d}{dt}\langle(\hat{x}-X_0)\Pi_x\rangle=0\), one obtains the identity
\(\langle \hat{x}\,\alpha_y\rangle=\ell_B^2\langle \Pi_x/\hbar \rangle\).
A short computation then yields the well-known result:
\begin{equation}
\label{eq:Sem_result}
\frac{S_{\rm EM}}{\hbar\omega_c T}=-(n+\tfrac{1}{2})\,.
\end{equation}

\paragraph{Spin contribution.}
If one isolates the spin-field coupling as a Pauli term
\(-\frac{e\hbar}{2m}\,\boldsymbol{\Sigma}\!\cdot\!\mathbf B\), its expectation
is constant for uniform \(B\):
\begin{equation}
\label{eq:Ssp_result}
S_{\rm SP}=\int_0^T\!\Big\langle -\frac{e\hbar}{2m}\,\boldsymbol{\Sigma}\!\cdot\!\mathbf B\Big\rangle dt
=-\,\frac{e\hbar B}{2m}\,s\,T
\;\;\Longrightarrow\;\;
\frac{S_{\rm SP}}{\hbar\omega_c T}=-\frac{s}{2}\,.
\end{equation}

\paragraph{Kinetic term.}
Define 
\(L_{\rm kin}=-mc^2\sqrt{w_\mu w^\mu}+mc^2\) with the drift fixed by the
Dirac eigenstate. On the Dirac mass shell one has the exact operator identity
\((\gamma_\mu\Pi^\mu)^2=\Pi_\mu\Pi^\mu-\tfrac{e\hbar}{2}\sigma^{\mu\nu}F_{\mu\nu}
= m^2c^2\) on \(|\Psi\rangle\), implying that the ``mechanical'' part of the action
is fully saturated by the EM and spin couplings. Consequently the kinetic contribution
makes no net addition over a stationary period:
\begin{equation}
\label{eq:Skin_zero}
S_{\rm kin}=0\,.
\end{equation}

\paragraph{Total Action.}
Summing \eqref{eq:Sem_result} and \eqref{eq:Ssp_result} with \eqref{eq:Skin_zero} gives:
\begin{equation}
\label{eq:Stot_result}
\frac{S_{\rm tot}}{\hbar\omega_c T}
=-(n+\tfrac{1}{2})-\frac{s}{2}
= -\Big(n+\frac{1+s}{2}\Big),
\end{equation}
which matches the phase \(-E_{n,s}T/\hbar\) extracted from the spectrum
\eqref{eq:LandauEnergy_SI} linearized in \(\hbar\omega_c\).
\paragraph{Remarks on gauge and conventions.}
The split \(S=S_{\rm kin}+S_{\rm EM}+S_{\rm SP}\) is convention-dependent:
\(S_{\rm EM}\) and \(S_{\rm SP}\) change under \(A_\mu\!\to\!A_\mu+\partial_\mu\chi\),
while their sum (and \(S_{\rm tot}\)) is gauge-invariant. In Landau gauge
the pieces take the simple forms above; all results quoted for \(S_{\rm tot}\)
are gauge independent.

\subsection{It\^o covariation counterterm for $S_{\rm EM}$ in the Dirac--Landau problem}
\label{app:landau-ito-half}

The physical electromagnetic action is obtained by integrating the local physical bilinear defined in Section~\ref{sec:ito-observables}.
For the charge-field term with general $A_\mu(z)$ and drift $w^\mu(z)$:
\begin{equation}
\label{eq:EM-phys-local}
  L_{\mathrm{EM}}^{\mathrm{phys}}(\tau)
  \;=\;
  e\,A_\mu\!\big(z_\tau\big)\,w^\mu\!\big(z_\tau\big)
  \;+\;\frac{e}{2}\sum_{\nu}(\sigma^\nu)^2\,
  \partial_\nu A_\mu\!\big(z_\tau\big)\,\partial_\nu w^\mu\!\big(z_\tau\big).
\end{equation}
Specializing this rule to Landau gauge at $k_y=0$, the covariation contribution is:
\begin{equation}
  \Delta L_{\mathrm{EM}}(\tau)
  \;=\; \frac{e}{2}\sum_{\nu}(\sigma^\nu)^2\,
  \partial_\nu A_\mu\!\big(z_\tau\big)\,\partial_\nu w^\mu\!\big(z_\tau\big).
\end{equation}
For $A_y=Bx$ and $w^y=\omega_c\,x$ only the $x$-channel contributes, with
$\partial_x A_y=B$ and $\partial_x w^y=\omega_c$, hence:
\begin{equation}
  \Delta L_{\mathrm{EM}}(\tau)
  \;=\; \frac{e}{2}\,(\sigma^x)^2\,B\,\omega_c .
  \label{eq:app-deltaEM-landau}
\end{equation}
The normalized action shift is:
\begin{equation}
  \frac{\Delta S_{\mathrm{EM}}}{\hbar\,\omega_c\,T}
  \;=\; \frac{\Delta L_{\mathrm{EM}}}{\hbar\,\omega_c}
  \;=\; \frac{e}{2\hbar}\,(\sigma^x)^2\,B .
  \label{eq:app-deltaS-norm}
\end{equation}
With the constant-diffusion choice $(\sigma^x)^2=\hbar/m$ and $\omega_c=|e|B/m$:
\begin{equation}
  \frac{\Delta S_{\mathrm{EM}}}{\hbar\,\omega_c\,T}
  \;=\; \frac{1}{2}.
\end{equation}
Thus the It\^o covariation adds a universal $+\tfrac12$ to the electromagnetic contribution in this normalization.

\subsection{It\^{o} nonlinear square-root prescription for $S_{\rm kin}$}
\label{app:kinetic-closure-eigenstate}

The kinetic contribution in the action is the nonlinear composition:
\begin{equation}
L_{\rm kin}(w)\;=\;-mc\,\sqrt{w_\mu w^\mu}\;+\;mc^2,
\label{eq:app_Lkin_nonlinear}
\end{equation}
cf.\ Eq.~\eqref{eq:Lsplit}. In the SOC evaluation we insert the optimal feedback drift $w^\mu=w^{\mu\star}(z)$ reconstructed from the Hopf-Cole field (Eq.~\eqref{eq:w-from-phi-tau}) and accumulate the running costs along It\^{o} trajectories.
Unlike the bilinear minimal-coupling term, $\sqrt{w^2}$ is a nonlinear function of a state-dependent field, so the physical-bilinear prescription of Section~\ref{sec:ito-observables} does not by itself determine a unique It\^{o} correction for $L_{\rm kin}$ without introducing an additional nonlinear calculus rule for composite observables.

The electromagnetic term is bilinear in the sense of Section~\ref{sec:ito-observables}, since it contains the local product $A_\mu(z)\,w^\mu(z)$; consequently its It\^{o} interpretation is fixed uniquely by Eq.~\eqref{eq:pi2-phys-local}. 
The corresponding corrected running cost $L_{\rm EM}^{\rm phys}$ is given explicitly in Appendix~\ref{app:landau-ito-half} (Eq.~\eqref{eq:EM-phys-local}).
In parallel, Section~\ref{sec:ito_cov} applies the same bilinear prescription to the invariant $\pi^2=\pi_\mu\pi^\mu$ (with $\pi_\mu:=m w_\mu$) and shows that, for Dirac energy eigenstates under the averaging assumptions stated there, the physical-bilinear mass shell satisfies Eq.~\eqref{eq:chi_mass_shell_avg}: the quantum-potential term entering the local mass shell is absorbed (in the mean) by the covariation part built into $(\pi^2)_{\rm phys}$, and the remaining correction beyond $m^2c^2$ is the Pauli invariant.
Since the Pauli invariant is already accounted for separately by the spin running cost $L_{\rm SP}$ in our split, we close the kinetic square-root on the corresponding spin-subtracted physical shell.
Within the present split this implies that the kinetic part carries no net phase contribution for the energy-eigenstate solutions under consideration:
\begin{equation}
S_{\rm kin}=0.
\label{eq:app_Skin_zero}
\end{equation}
Operationally, Eq.~\eqref{eq:app_Skin_zero} should be viewed as an on-shell normalization for the nonlinear square-root term: it fixes the finite ambiguity of the kinetic component that remains once the bilinear It\^{o} prescription has been applied to the minimal-coupling term.

With \eqref{eq:app_Skin_zero}, the total action is reconstructed from the terms whose It\^{o} interpretation is already fixed within the manuscript:
\begin{equation}
S_{\rm tot}^{\rm phys}
=\int_0^T\!\Big(L_{\rm EM}^{\rm phys}(z_\tau,w^\star(z_\tau)) + L_{\rm SP}\Big)\,d\tau,
\label{eq:app_Stot_reconstruct}
\end{equation}
where $L_{\rm EM}^{\rm phys}$ is taken from Appendix~\ref{app:landau-ito-half} and $L_{\rm SP}$ is the deterministic spin (see Eq.~\eqref{eq:Lsplit}).
Equation~\eqref{eq:app_Stot_reconstruct} is the prescription used in the benchmark comparisons: the bilinear EM contribution is evaluated with the unique covariation correction, the Pauli/spin contribution is included explicitly, and the kinetic square-root is closed on shell via \eqref{eq:app_Skin_zero}.

\section{Complex quantum mass-shell geometry}
\label{sec:landau-mass-shell}

In the Complex SOC theory the kinetic four-momentum $\pi_\mu(z)$ defined in Eq.~\eqref{eq:pi_momentum} is a complex field. 
The quantum mass-shell relation:
\begin{equation}
\label{eq:mass_shell_S}
  \pi_\mu \pi^\mu = S(z)
\end{equation}
involves the complex scalar $S(z)$ defined in Eq.~\eqref{eq:S_scalar_def}.

To make the complex structure explicit we decompose the momentum as:
\begin{equation}
  \pi_\mu = p_\mu + i q_\mu,
  \qquad
  p_\mu := \Re \pi_\mu,
  \quad
  q_\mu := \Im \pi_\mu .
\label{eq:pq-split}
\end{equation}
Writing $S(z) = S_{\mathrm{R}}(z) + i\, S_{\mathrm{I}}(z)$ with real and imaginary parts $S_{\mathrm{R}}, S_{\mathrm{I}}$ and inserting Eq.~\eqref{eq:pq-split} into Eq.~\eqref{eq:mass_shell_S} we obtain the pair of real relations:
\begin{align}
  p_\mu p^\mu - q_\mu q^\mu &= S_{\mathrm{R}}(z),
  \label{eq:pq-real-part} \\
  2\, p_\mu  q^\mu &= S_{\mathrm{I}}(z).
  \label{eq:pq-imag-part}
\end{align}
Equation~\eqref{eq:pq-real-part} can be viewed as a shifted shell relation for the real part of the momentum:
\begin{equation}
  p_\mu p^\mu
  = S_{\mathrm{R}}(z) + q_\mu q^\mu.
\label{eq:p-shifted-shell}
\end{equation}
In the Complex SOC formulation the constraint \eqref{eq:mass_shell_S} defines a six-dimensional subset of the full eight-dimensional $(p_\mu,q_\mu)$ space. Its projection onto the four-dimensional real momentum space $\mathbb{R}^4$ cannot, in general, be represented by a single three-dimensional real hypersurface.
Moreover, because $S(z)$ is state- and position-dependent and is sampled along stochastic worldlines $z(\tau)$, an ``exact'' visualization of the quantum mass shell in real momentum space is, in general, only accessible in a trajectory-based manner by plotting the sampled points $p_\mu(\tau)=\Re\pi_\mu\bigl(z(\tau)\bigr)$ rather than by drawing a closed surface.

Within real momentum space, it is therefore convenient to introduce a kinematic reference family labelled by the magnitude of the imaginary spatial momentum.
Writing $\pi_{\mathrm{sp}} := (\pi_x,\pi_y,\pi_z)$ for the spatial part of $\pi_\mu$, we define:
\begin{equation}
  \varrho \;\equiv\; \frac{\|\Im \pi_{\mathrm{sp}}\|}{mc} .
\label{eq:rho-def}
\end{equation}

We then construct the \emph{weak mass shell} by fixing the imaginary spatial deformation $\varrho$ and enforcing the weak closure $\pi_\mu\pi^\mu=m^2c^2$ in place of the full quantum mass-shell factor $S(z)$ in the defining constraint.
We therefore define:
\begin{equation}
  \widetilde{\mathcal{C}}_\varrho := \big\{ \pi \in \mathbb{C}^4 :
  \pi_\mu \pi^\mu = m^2 c^2,\,
  \|\Im \pi_{\mathrm{sp}}\|/(mc) = \varrho \big\},
\label{eq:C-rho-tilde}
\end{equation}
and its real projection:
\begin{equation}
  \widetilde S_\varrho := \{ p = \Re \pi : \pi \in \widetilde{\mathcal{C}}_\varrho \}
  \subset \mathbb{R}^4.
\label{eq:S-rho-tilde}
\end{equation}
Because the weak closure in \eqref{eq:C-rho-tilde} has a purely real right-hand side, the imaginary-part relation \eqref{eq:pq-imag-part} reduces on $\widetilde C_{\varrho}$ to $p_\mu q^\mu=0$, i.e. Minkowski orthogonality between the real and imaginary four-momenta. 
This constraint couples the admissible time-like component $\Im\pi_0$ (hence $\Re\pi_0$) to the chosen spatial orientation at fixed $\|\Im\pi_{\mathrm{sp}}\|/(mc)=\varrho$.

The deviation of the exact quantum constraint~\eqref{eq:mass_shell_S} from the weak closure $\pi_\mu\pi^\mu=m^2c^2$ is entirely due to the quantum corrections encoded in $S(z)$ (quantum-potential and Pauli terms).
A practical motivation for introducing $\widetilde S_\varrho$ is that it provides a purely kinematic reference surface: it is obtained by enforcing the weak closure $\pi_\mu\pi^\mu=m^2c^2$ at fixed $\varrho$, and therefore can be constructed without evaluating the full quantum mass-shell factor $S(z)$.
When the quantum corrections in $S(z)$ are small, $\widetilde S_\varrho$ provides a useful baseline and the projected trajectory points $p_\mu(\tau)$ can be interpreted as fluctuating around the corresponding weak shell at the deformation level set by $\varrho(\tau)$.

For the Dirac-Landau problem considered here, these additional contributions in $S(z)$ are small compared to $m^2c^2$ and to the geometric deformation set by $\varrho$, as evidenced by the tiny mass-shell residual in
Section~\ref{subsec:mass-shell-residual}. 
This makes it natural to work with the weak mass shells $\widetilde S_\varrho$, which retain the imaginary spatial deformation while treating the remaining quantum corrections as small trajectory-level deviations rather than as a change in the plotted shell geometry itself.

For reference we denote by $S_0$ the classical mass shell in real momentum space:
\begin{equation}
  S_0 = \big\{ p \in \mathbb{R}^4 : p_\mu p^\mu = m^2 c^2 \big\},
\label{eq:S0-momentum}
\end{equation}
which coincides with the classical Hamilton-Jacobi mass shell. 

In the Complex SOC setting the sets $\widetilde S_\varrho$ form a $\varrho$-labelled family of real reference shells. 
In the WKB classical limit $\hbar\to 0$ one has $\Im \pi_\mu \to 0$, hence $\varrho \to 0$, and the family $\widetilde S_\varrho$ collapses onto the classical shell $S_0$.
In the same limit the projected momentum trajectories $p_\mu(\tau)=\Re\pi_\mu(\tau)$ also collapse onto $S_0$, i.e. $p_\mu(\tau)p^\mu(\tau)\to m^2c^2$, recovering the classical Hamilton-Jacobi mass shell pointwise along the paths.
For $\varrho > 0$ the sets $\widetilde S_\varrho$ provide reference shells for real projections of genuinely complex on-shell momenta with nonzero imaginary spatial components.

\begin{figure}[ht!]
  \centering
    \vspace{-10pt}
  \includegraphics[width=0.85\linewidth]{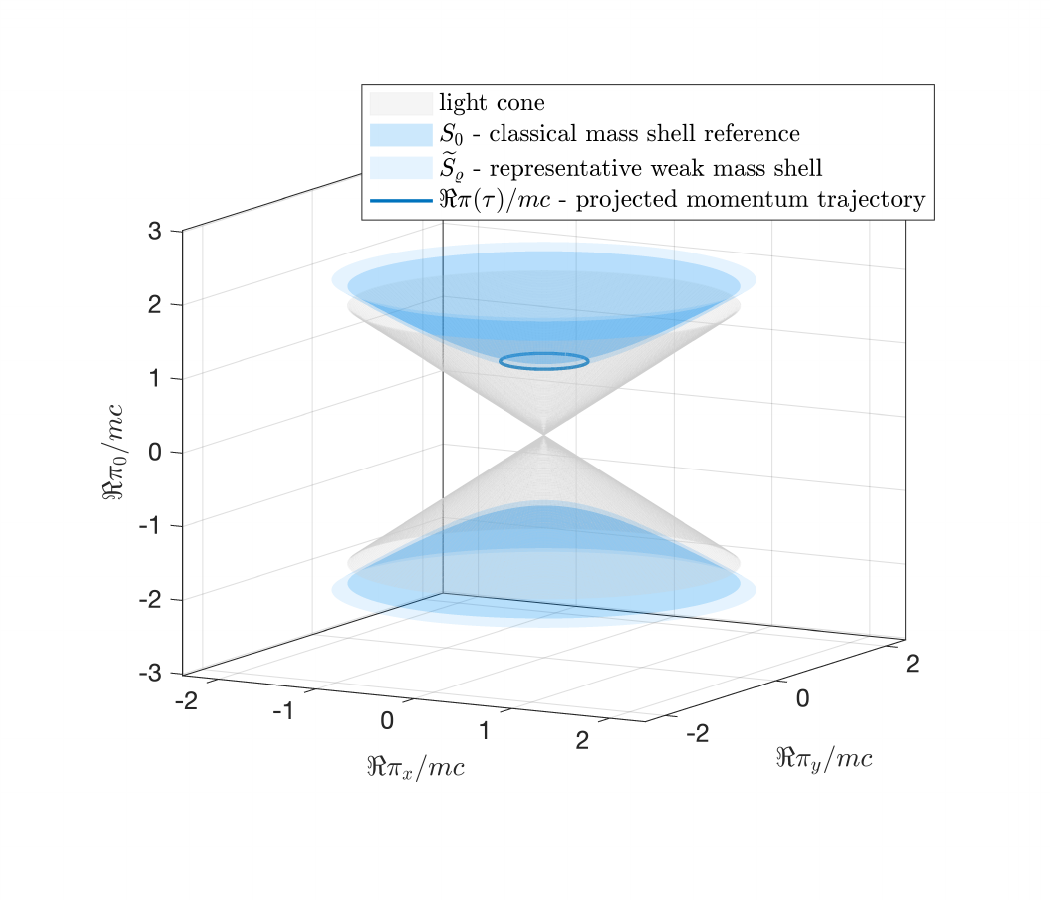}
    \vspace{-30pt}
  \caption{Complex momentum mass-shell geometry obtained from the Complex SOC numerical simulation of the Dirac-Landau problem (here $n=1$, $s=-1$, $B=0.1\,\mathrm{T}, k_y= 1 \times 10^{12}\, \mathrm{m}^{-1}$) in the slice $p_z=0$. 
  The light cone $p_0^2 = p_x^2 + p_y^2$ (grey) and the classical Hamilton-Jacobi shell $S_0$ (blue) are shown together with a representative weak mass shell $\widetilde S_{\varrho_\text{eff}}$ (light blue), where $\varrho_\text{eff}$ is the RMS level of the imaginary spatial momentum, $\varrho^2_\text{eff}=\left<\varrho(\tau)^2\right>_\tau$, measured along the Complex SOC Dirac-Landau trajectory. 
  The dark blue curve depicts the real projection $p_\mu(\tau)/(mc) = \Re\pi_\mu(\tau)/(mc)$ of representative Complex SOC trajectory generated by the stochastic algorithm.}
  \label{fig:mass-shell-geometry}
\end{figure}

To visualize this geometry we work with three-dimensional slices of real momentum space. 
For concreteness we fix $p_z=0$ and introduce the dimensionless coordinates:
\begin{equation}
\label{eq:real_mass_shell_slice}
  \Bigl( \frac{p_x}{mc}, \frac{p_y}{mc}, \frac{p_0}{mc} \Bigr),
\end{equation}
which provide the axes in figure~\ref{fig:mass-shell-geometry}.

In the Complex SOC numerical simulation of the Dirac-Landau problem from Section~\ref{sec:numerics} we obtain sample paths of the complex momentum $\pi_\mu(\tau) = \pi_\mu\bigl(z(\tau)\bigr)$ as a function of proper time $\tau$.
Along these paths we compute:
\begin{equation}
  \varrho(\tau) = \frac{\|\Im \pi_{\mathrm{sp}}(\tau)\|}{mc},
\end{equation}
and from these values define an effective level of imaginary spatial momentum via $\varrho^2_\text{eff} = \langle \varrho(\tau)^2\rangle_\tau$.
We then construct a representative surface $\widetilde S_{\varrho_\text{eff}}$ in the slice defined by Eq.~\eqref{eq:real_mass_shell_slice}. 
For visualization we render a representative $\widetilde S_{\varrho}$ by choosing the typical relative orientation between $\Re\pi_{\mathrm{sp}}$ and $\Im\pi_{\mathrm{sp}}$ observed along the simulated trajectory at the same deformation level.
On the same plot, we superimpose the projected momentum trajectory $p_\mu(\tau)/(mc) = \Re\pi_\mu(\tau)/(mc)$.

This yields the geometry illustrated in figure~\ref{fig:mass-shell-geometry}: a reference light cone, the classical shell $S_0$, a representative deformed shell $\widetilde S_{\varrho_\text{eff}}$ corresponding to the RMS level of imaginary spatial momentum along the trajectory, and the paths traced by the real part of the complex momentum. 
In the Dirac-Landau example shown, the trajectory points lie very close to $\widetilde S_{\varrho_\text{eff}}$, with the small residual deviations quantified in Section~\ref{subsec:mass-shell-residual}.

The parameter choices in figures~\ref{fig:mass-shell-geometry} and ~\ref{fig:mass-shell-residual} are intentionally different: figure~\ref{fig:mass-shell-geometry}  uses a large $k_y$ to improve geometric visibility of the light cone relative to the mass-shell surface at the plotted scale, whereas figure~\ref{fig:mass-shell-residual} is chosen to make the small stochastic fluctuations about the smooth momentum evolution clearly visible.
 
\subsection{Deviation of the quantum mass shell from the weak mass shell}
\label{subsec:mass-shell-residual}

The quantum mass-shell relation Eq.~\eqref{eq:mass_shell_S} implies that, in general, the kinetic four-momentum $\pi_\mu$ does not satisfy the weak closure $\pi_\mu\pi^\mu=m^2c^2$ exactly, but is shifted by the quantum potential and Pauli contributions collected in $S(z)$, cf.~Eq.~\eqref{eq:S_scalar_def}.
In the Complex SOC theory, the weak closure $\pi_\mu\pi^\mu=m^2c^2$ can be viewed as the direct complex extension of the classical mass-shell constraint, since $\pi_\mu(z)$ is complex-valued along the evolution.
The quantum mass shell Eq.~\eqref{eq:mass_shell_S} then represents a further deformation of this closure through the
replacement of $m^2c^2$ with $S(z)$.
Accordingly, the weak mass shell provides a convenient reference geometry for quantifying the size of these additional contributions in the Dirac-Landau simulations considered here.

It is nevertheless useful to quantify, for a given Complex SOC trajectory, how closely the kinetic momentum satisfies the weak closure $\pi_\mu\pi^\mu=m^2c^2$.
To this end we define the dimensionless residual:
\begin{equation}
  \delta(\tau)
  := \frac{\pi_\mu(\tau)\,\pi^\mu(\tau) - m^2 c^2}{m^2 c^2},
\end{equation}
and monitor its real and imaginary parts along the simulated paths.

\begin{figure}[ht!]
  \centering
  \includegraphics[width=0.7\linewidth]{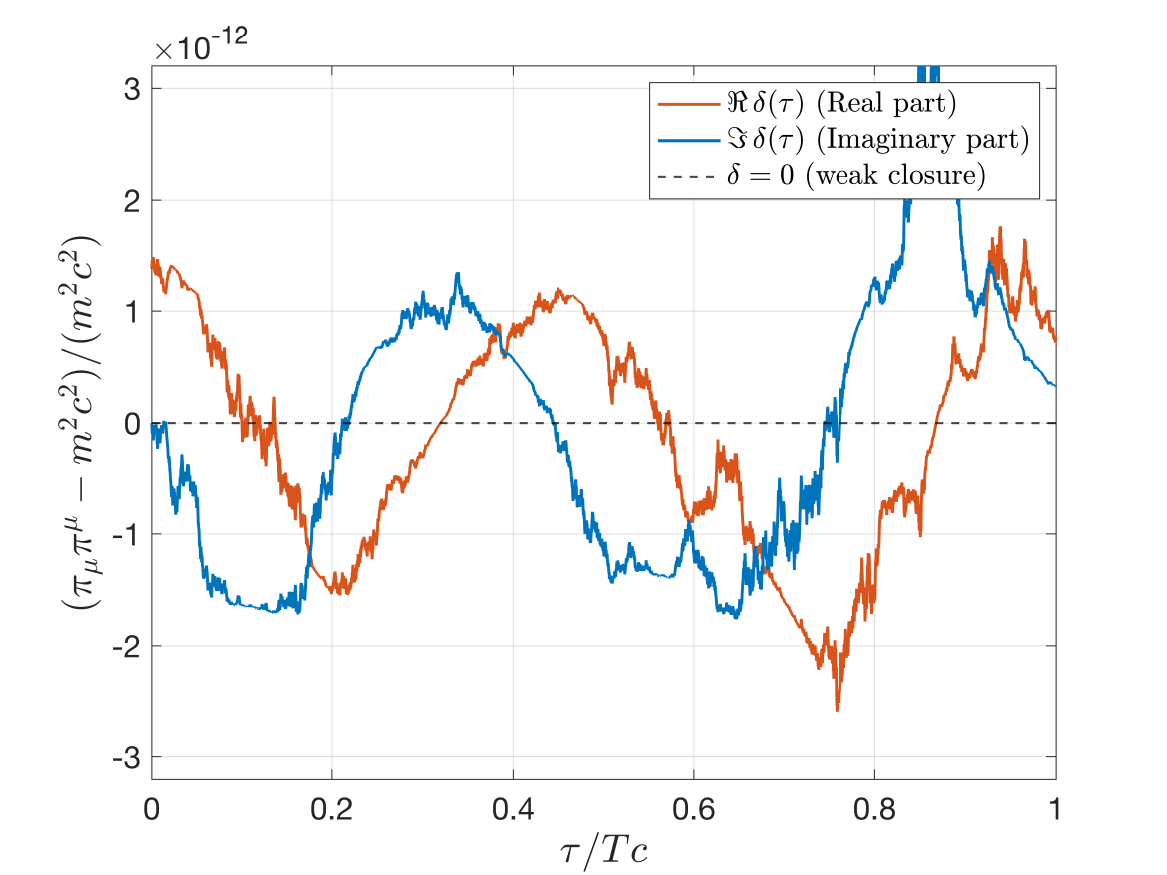}
  \caption{Deviation of the kinetic four-momentum from the weak closure for a Complex SOC electron trajectory in the Dirac-Landau problem.
Plotted is the dimensionless residual $\delta(\tau)=\bigl[\pi_\mu(\tau)\pi^\mu(\tau)-m^2 c^2\bigr]/(m^2 c^2)$ as a function of dimensionless time $\tau/T_c$, for $n=1$, $s=-1$, $B=0.1\,\mathrm{T}$, $k_y = 5 \times 10^{7}\,\mathrm{m}^{-1}$ and $p_z=0$.
The real and imaginary parts remain at the level of a few $\times 10^{-12}$ and exhibit bounded oscillations with period close to $T_c/2$, approximately $\pi/2$ out of phase.
This single-trajectory diagnostic shows that the quantum-potential and Pauli contributions encoded in $S(z)$ induce only a very small complex deformation of the weak closure $\pi_\mu\pi^\mu=m^2c^2$ in the present parameter regime.}
  \label{fig:mass-shell-residual}
\end{figure}

Figure~\ref{fig:mass-shell-residual} shows $\Re\delta(\tau)$ and $\Im\delta(\tau)$ along a representative Complex SOC electron trajectory in the Dirac-Landau setting, evolved over one cyclotron period.
The trajectory in momentum space is stochastic due to the underlying complex drift-diffusion dynamics, but for the selected parameters the deterministic drift dominates and the random fluctuations appear as small oscillatory deviations around an underlying smooth signal. 
Both components remain bounded and oscillatory, with amplitudes of order $10^{-12}$ and no secular drift over many time steps. 
The oscillations have a period close to $T_c/2$ and the real and imaginary parts are approximately $\pi/2$ out of phase, which is consistent with the analytic structure of the scalarised mass shell for the $n=1$ Landau level: along a nearly circular cyclotron orbit the relevant quantum corrections depend on the square of a complex phase factor, so that the residual acquires a dominant harmonic at frequency $2\omega_c$ with real and imaginary parts in quadrature.

Overall, the figure confirms that the quantum mass shell defined by Eq.~\eqref{eq:mass_shell_S} remains very close to the weak mass shell $\pi_\mu \pi^\mu = m^2 c^2$, while the small complex deviations encode the quantum corrections discussed in Section~\ref{sec:landau-mass-shell}.

\section{Visualization of coordinate-space worldlines, light cones, and causal residuals}
\label{sec:coord_lightcone}

The complex mass-shell geometry analysed in Appendix~\ref{sec:landau-mass-shell} is formulated in momentum space.
Here we present a complementary coordinate-space visualization based on the numerical Complex SOC solution of the
Dirac-Landau problem. Using a representative simulated trajectory $z^\mu(\tau)$ in the $p_z=0$ Landau slice, we display the associated real-projected worldline in space-time together with light-cone diagnostics that quantify its position relative to the Minkowski cone.

In Complex SOC theory the particle worldline is represented by a complex space-time coordinate $z^\mu(\tau)\in\mathbb C^4$. For geometric visualization in coordinate space we use the real projection $\Re z^\mu(\tau)$.
In the Dirac-Landau setting with $\mathbf B\parallel \hat{\mathbf z}$ and in the $p_z=0$ slice used in the preceding figures, the transverse plane $(x,y)$ is the natural spatial projection.
Throughout this section we therefore set:
\begin{equation}
  ct(\tau)\equiv \Re z^0(\tau),\qquad
  x(\tau)\equiv \Re z^1(\tau),\qquad
  y(\tau)\equiv \Re z^2(\tau),
  \label{eq:ctxy_def}
\end{equation}
where $ct$ has the dimension of length. In the simulations underlying the figures we choose $z^\mu(0)=0$, so that the cone apex coincides with the initial event.

Define the transverse radius:
\begin{equation}
  r_\perp(\tau)\equiv \sqrt{x(\tau)^2+y(\tau)^2}.
  \label{eq:rperp_def}
\end{equation}
In the $(x,y)$-projection the light-cone boundary is:
\begin{equation}
  ct = r_\perp,
  \label{eq:cone_boundary}
\end{equation}
and the associated projected causal interval (in the $(x,y)$-section) is:
\begin{equation}
  s_\perp^2 \equiv (ct)^2-r_\perp^2 .
  \label{eq:sperp_def}
\end{equation}
Evaluating these expressions along the worldline yields the scalar diagnostic $s_\perp^2(\tau)=(ct(\tau))^2-r_\perp(\tau)^2$.
With metric signature $(+,-,-,-)$, the condition $s_\perp^2\ge 0$ corresponds to lying inside the projected cone, while $s_\perp^2=0$ corresponds to the boundary.

In parameter regimes where the deterministic trend dominates the raw worldline, stochastic fluctuations may be visually subtle on the full coordinate-space scale. To isolate the stochastic contribution without modifying physical parameters, we subtract the drift-only reconstruction obtained by integrating the simulated drift:
\begin{equation}
  z^{\mu}_{\mathrm{drift}}(\tau_{k+1})
  \,=\, z^{\mu}_{\mathrm{drift}}(\tau_k) + \Re w^\mu(\tau_k)\,\Delta\tau,
  \qquad z^{\mu}_{\mathrm{drift}}(0)=0,
  \label{eq:drift_recon}
\end{equation}
and define:
\begin{equation}
ct_{\mathrm{drift}}(\tau)\equiv \Re z^0_{\mathrm{drift}}(\tau), \qquad r_{\perp,\mathrm{drift}}(\tau)\equiv \sqrt{x_{\mathrm{drift}}(\tau)^2+y_{\mathrm{drift}}(\tau)^2}.
\end{equation}
In Eq.~\eqref{eq:drift_recon} the drift samples $w^\mu(\tau_k)$ are those produced by the same Euler-Maruyama realization, i.e. $w^\mu(\tau_k)=w^\mu(z(\tau_k))$, and are stored during the stochastic simulation. Thus $z^\mu_{\rm drift}$ is a post-processed cumulative drift contribution for that realization; we do not re-evaluate $w^\mu$ along $z_{\rm drift}$ as in a separately propagated deterministic ODE.

Applying Eq.~\eqref{eq:sperp_def} to the drift-only path yields $s_{\perp,\mathrm{drift}}^2(\tau)$.
We then form the projected causal residual about drift:
\begin{equation}
  \delta s_\perp^2(\tau)\equiv s_\perp^2(\tau)-s_{\perp,\mathrm{drift}}^2(\tau),
  \label{eq:causal_residual}
\end{equation}
which provides a compact scalar measure of coordinate-space stochasticity relative to the light-cone geometry in the chosen projection.

\begin{figure}[ht!]
  \centering
      \vspace{-10pt}
     \includegraphics[width=0.6\linewidth]{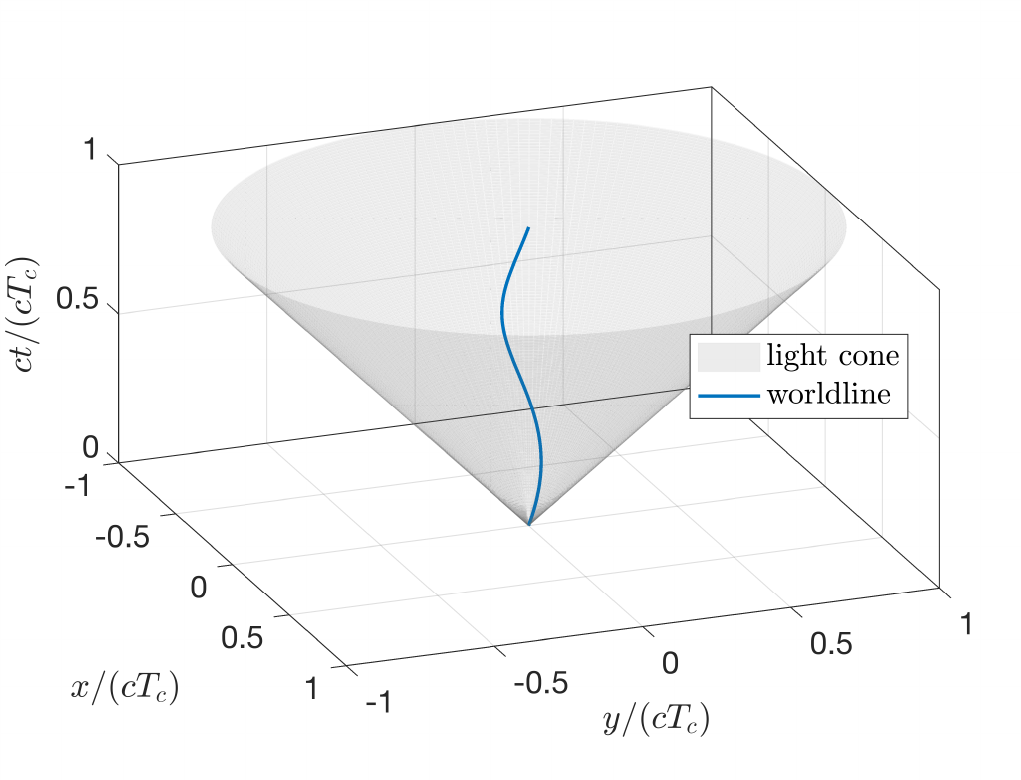}
   \vspace{-10pt}
   \caption{Coordinate-space light cone and representative worldline in the Dirac--Landau simulation.
Plotted is the real-projected worldline $(x(\tau),y(\tau),ct(\tau))$ together with the projected light-cone surface $ct=\sqrt{x^2+y^2}$. 
   Simulation parameters: $n=1$, $s=-1$, $B=0.1\,\mathrm{T}$, $k_y=10^{12}\,\mathrm{m^{-1}}$, $p_z=0$, and $z^\mu(0)=0$.}
  \label{fig:worldline_cone_3D}
\end{figure}

Figure~\ref{fig:worldline_cone_3D} shows a representative worldline in $(x,y,ct)$ together with the projected light-cone surface $ct=\sqrt{x^2+y^2}$.

\begin{figure}[ht!]
  \centering
  \includegraphics[width=0.49\linewidth]{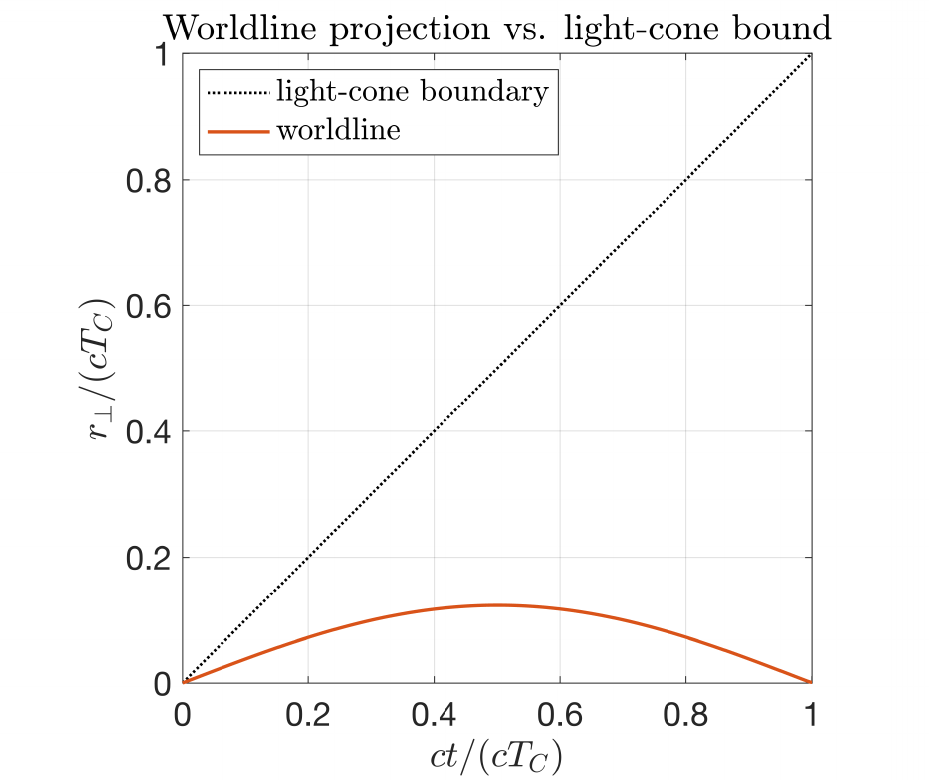}\hfill
  \includegraphics[width=0.49\linewidth]{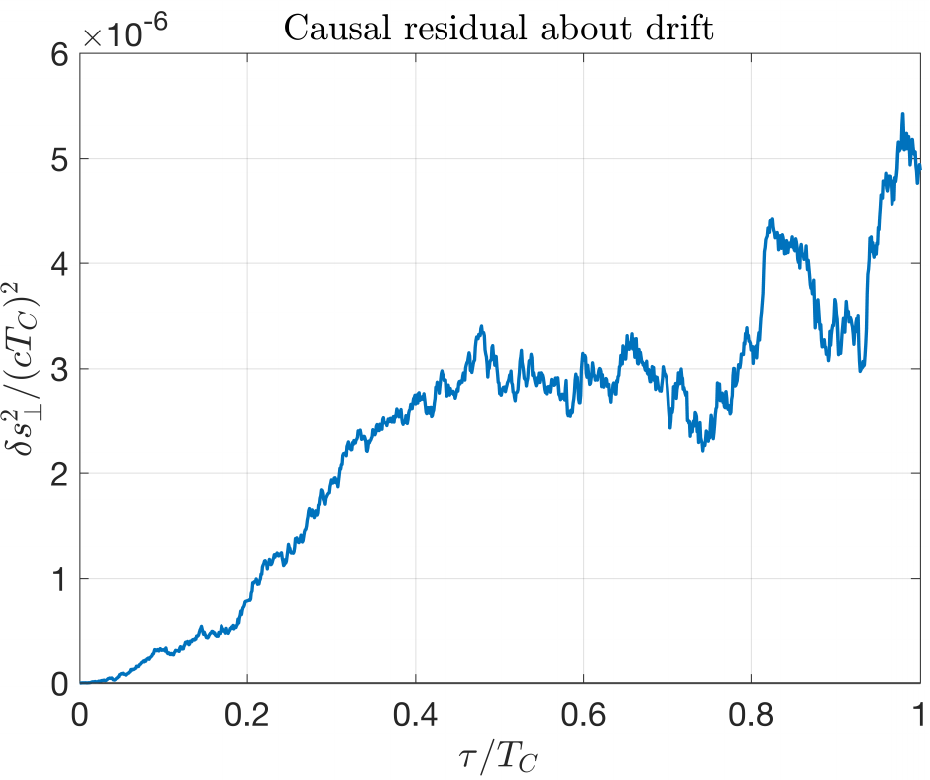}
  \caption{Coordinate-space light-cone diagnostics for the same trajectory as Fig.~\ref{fig:worldline_cone_3D}. \textbf{Left}: spacetime projection $r_\perp(\tau)$ versus $ct(\tau)$; with the light-cone boundary $r_\perp=ct$. \textbf{Right}: projected causal residual about drift $\delta s_\perp^2(\tau)\equiv s_\perp^2(\tau)-s_{\perp,\mathrm{drift}}^2(\tau)$
  normalized by $(cT_c)^2$, where $s_\perp^2=(ct)^2-r_\perp^2$ is given by Eq.~\eqref{eq:sperp_def} and the drift-only reconstruction
  is defined by Eq.~\eqref{eq:drift_recon}.
  Simulation parameters are the same as in figure~\ref{fig:worldline_cone_3D}.}
  \label{fig:cone2D_and_residual}
\end{figure}

Figure~\ref{fig:cone2D_and_residual} (left) displays the corresponding spacetime projection $r_\perp$ versus $ct$, with the cone boundary given by $r_\perp=ct$, while the right panel shows the projected causal residual about drift $\delta s_\perp^2(\tau)$ along the same trajectory.

Taken together, these plots provide a direct coordinate-space visualization of the Complex SOC diffusion, and the residual isolates the stochastic contribution relative to the deterministic drift for the same parameter set.

\bibliographystyle{unsrt} 
\bibliography{refs}

\section*{Data Availability}
All code and inputs required to reproduce the figures and table are available in
the repository cited as Ref.~\cite{Yordanov2025Code}. Step-by-step
instructions to run the MATLAB scripts and regenerate Figures~1-2 and the
action-components table are provided in the repository \texttt{README.md}.
No external datasets were used; all outputs are generated by the provided simulation code.

\end{document}